\documentclass[fleqn,usenatbib]{mnras}

\usepackage{newtxtext,newtxmath}

\usepackage[T1]{fontenc}
\usepackage{ae,aecompl}
\usepackage[usenames, dvipsnames]{color}

\usepackage{graphicx}	
\usepackage{amsmath}	
\usepackage{amssymb}
\usepackage{multirow}
\newcommand{\blue}[1]{{\color{black}{#1}}}
\newcommand{\red}[1]{{\color{black}{#1}}}

\title[Soft excess in RX J0812.4-3114]{Soft excess in the quiescent Be/X-ray pulsar RX J0812.4-3114}

\author[Yue Zhao et al.]{
Yue Zhao,$^{1}$\thanks{E-mail: zhao13@ualberta.ca}
Craig O. Heinke,$^{1}$
Sergey S. Tsygankov,$^{2,3}$
Wynn C. G. Ho,$^{4,5}$
\newauthor
Alexander Y. Potekhin,$^{6}$
and
Aarran W. Shaw$^{1}$
\\
$^{1}$ Department of Physics, University of Alberta, CCIS 4-183, Edmonton, AB T6G 2E1, Canada\\
$^{2}$ Department of Physics and Astronomy, FI-20014 University of Turku, Finland\\
$^{3}$ Space Research Institute of the Russian Academy of Sciences, Profsoyuznaya Str. 84/32, Moscow 117997, Russia\\
$^{4}$ Department of Physics and Astronomy, Haverford College, 370 Lancaster Avenue, Haverford, PA 19041, USA \\
$^{5}$ Mathematical Sciences, Physics and Astronomy, and STAG Research Centre, University of Southampton, Southampton SO17 1BJ, UK \\
$^{6}$ Ioffe Institute, Politekhnicheskaya 26, 194021, Saint Petersburg, Russia\\
}

\date{Accepted XXX. Received YYY; in original form ZZZ}

\pubyear{2018}

\begin{document}
\label{firstpage}
\pagerange{\pageref{firstpage}--\pageref{lastpage}}
\maketitle

\begin{abstract}
We report a 72 ks {\it XMM-Newton} observation of the Be/X-ray pulsar (BeXRP) RX J0812.4-3114 in quiescence ($L_X \approx 1.6\times 10^{33}~\mathrm{erg~s^{-1}}$). 
Intriguingly, we find a two-component spectrum, with a hard power-law ($\Gamma\approx 1.5$) and a soft blackbody-like excess below $\approx 1$ keV. 
The blackbody component is consistent in $kT$ with a prior quiescent {\it Chandra} observation reported by Tsygankov et al. and has an inferred blackbody radius of $\approx$ 10 km, consistent with emission from the entire neutron star (NS) surface.  
There is also mild evidence for an absorption line at $\approx 1~\mathrm{keV}$ \red{and/or $\approx 1.4~\mathrm{keV}$}. The hard component shows pulsations at $P\approx 31.908$ s (pulsed fraction $0.84\pm 0.10$), agreeing with the pulse period seen previously in outbursts, but no pulsations were found in the soft excess (pulsed fraction $\lesssim 31\%$). 
We conclude that the pulsed hard component suggests low-level accretion onto the neutron star poles, while the soft excess seems to originate from the entire NS surface.
We speculate that, in quiescence, the source switches between a soft thermal-dominated state (when the propeller effect is at work) and a relatively hard state with low-level accretion, and use the propeller cutoff to estimate the magnetic field of the system to be $\lesssim 8.4\times 10^{11}~\mathrm{G}$.
We compare the quiescent thermal $L_X$ predicted by the standard deep crustal heating model to our observations and find that RX J0812.4-3114 has a high thermal $L_X$, at or above the prediction for minimum cooling mechanisms. This suggests that RX J0812.4-3114 either contains a relatively low-mass NS with minimum cooling, or that the system may be young enough that the NS has not fully cooled from the supernova explosion.

\end{abstract}




\section{Introduction}
\label{sec:sec_intro}
Be/X-ray pulsars (BeXRPs) are a type of high-mass X-ray binary (HMXB) where a highly magnetised neutron star (NS; $B\sim 10^{11-13}$ G) regularly passes through the decretion disk expelled by a Be-type optical companion (for a recent review on Be stars, see \citealt{rivinius2013}). 
These systems are typically identified by their bright type-I outbursts ($L_X \sim 10^{36-37}~\mathrm{erg~s^{-1}}$;  \citealt{reig2011}) that happen during orbital periastron passages, when the NS ploughs through the decretion disc around the Be star, leading to a sharp increase in mass accretion rate.  The high NS magnetic fields channel accreted matter onto the magnetic poles, producing hard X-ray emission that pulses at the NS spin period.   Inflowing ionized matter is forced to move along magnetic field lines when the magnetic pressure equals the ram pressure of the infalling matter, defining the magnetospheric radius within which a hot disc will be disrupted. High magnetic fields in systems with short spin periods may halt or largely suppress accretion by forming centrifugal barriers when the magnetospheric radius is larger than the corotation radius (where the orbital period in the disc matches the NS spin period). In this situation, material threading onto magnetic field lines must be accelerated to higher velocities, which moves it outward through the disc, inhibiting accretion; this is known as the ``propeller regime" \citep{illarionov1975}.
As the mass accretion rate falls during an outburst decline, allowing the magnetosphere to expand, 
an abrupt drop in X-ray luminosity has  often been observed, suggesting the system has entered the propeller regime  \citep{stella86,campana01,tsygankov2016}. 

However,  signs of continued accretion are still observed in some low-luminosity systems that have seemingly transitioned to the propeller regime. E.g., pulsations from 3A 0535+26 were detected when the source was at  $L_X\approx 2-4\times 10^{33}~\mathrm{erg~s^{-1}}$  \citep{negueruela2000,mukherjee2005}. Multiple scenarios have been proposed to explain the mechanisms of matter penetrating the centrifugal barrier (e.g. \citealt{romanova2004,Doroshenko11}). \cite{tsygankov2017a} recently proposed that in systems with sufficiently long spin period, below a certain accretion rate, the disc temperature may fall below the hydrogen ionisation temperature ($\sim 6500$ K) rendering a recombined neutral ``cold disc'',
which can penetrate through the centrifugal barrier of the magnetosphere. Several sources have been observed to maintain quasi-stable accretion at an intermediate luminosity, higher than the limiting luminosity for the propeller regime (e.g., \citealt{rouco2018}). 

X-ray studies of BeXRPs in quiescence ($L_X<10^{34}~\mathrm{erg~s^{-1}}$) have revealed  hard X-ray spectra, typically best described by power-laws, suggesting continued accretion \citep[e.g.][]{campana2002,Rutledge07,Doroshenko14}, and/or soft blackbody-like spectra suggestive of emission from part of the NS surface \citep{LaPalombara06,LaPalombara07,LaPalombara09,Reig14,elshamouty2016}.
In a recent systematic study of quiescent BeXRPs by \cite{tsygankov2017a}, X-ray spectra of quiescent BeXRPs generally showed either  a soft blackbody-like spectrum with an emission region consistent with a typical NS polar cap size (e.g., 4U 0115+63, with $kT_\mathrm{bb}\approx 0.3~\mathrm{keV}$ and $R_\mathrm{bb}\approx 0.76~\mathrm{km}$), 
or a hard power-law spectral component (photon index $\Gamma$ typically $\sim$ 1-1.5) suggesting an accretion flow (e.g., 4U 0728-25 with $\Gamma\approx 1.3$). These suggest quiescent states either with (hard) or without (soft) continued accretion.

\begin{figure*}
    \centering
    \includegraphics[scale=0.6]{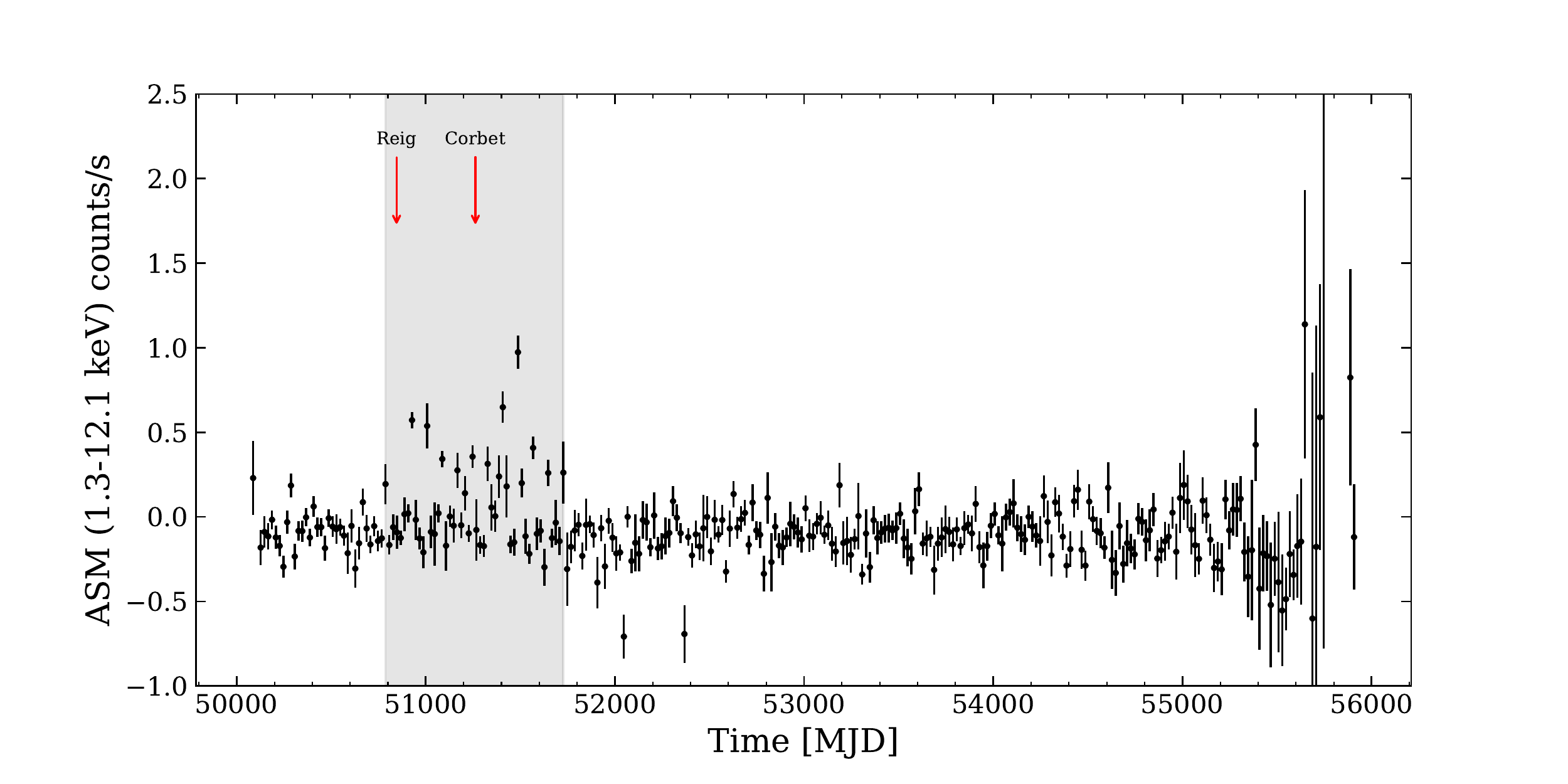}
    \caption{{\it RXTE}/ASM $1.3-12.1~\mathrm{keV}$ light curve of RX J0812.4-3114, rebinned to 10 days. The red arrows indicate two {\it RXTE}/PCA observations during outburst (two in Feb. 1998 marked with one arrow), and the grey shaded region approximately marks the active period of the source. The uncertainties increase near the end of {\it RXTE}'s mission in 2011, which was before the {\it Chandra} and {\it XMM-Newton} observations described here.}
    \label{fig:asm_lc}
\end{figure*}

RX J0812.4-3114 (hereafter J0812) was first identified by the ROSAT Galactic Plane Survey \citep{motch1991} as an X-ray source that positionally coincides with a Be star (LS 992, B0.2IVe; see \citealt{motch1997,reig2000}).   \cite{corbet1999} reported that the source entered an active state in early 1998, with a series of prominent outbursts. \citet{reig1999} reported two observations in February 1998 with the Proportional Counter Array (PCA) on the {\it Rossi X-ray Timing Explorer} ({\it RXTE}), with which they detected strong X-ray pulsations at a period of 31.885 s. The X-ray pulsar (XRP) nature of J0812 was hence corroborated. Using data from the All Sky Camera (ASM) on board RXTE, the orbital period was then found to be $\approx 81$ days by \cite{corbet2000}, who used the {\it RXTE}/PCA observation on March 25, 1999 to again confirm the strong pulsations at a period of $31.88$ s.  
Fig. \ref{fig:asm_lc} shows the full ASM light curve, from which we see that the source stayed in an active state until early July of 2000, and returned to a relatively low-count state ever since.

The {\it Chandra X-ray Observatory} observed J0812 in July 2013, as part of a campaign to systematically study quiescent BeXRPs (PI: Wijnands, ObsIDs: 14635-14650), during which it had a relatively low $L_X$ of $\sim 2 \times 10^{33}~\mathrm{erg~s^{-1}}$ and a very soft spectrum. A fit with an absorbed power-law gave a photon index $\Gamma$ of $\approx 5.6$, which suggests a blackbody-like fit would be more appropriate. Intriguingly, the blackbody fit gave an unusually low temperature of $kT\approx 0.1$ keV, suggesting thermal emission from a large but poorly constrained inferred emission radius (up to $\sim 10$ km; see \citealt{tsygankov2017b} for more details). 

In this work, we report results from our recent {\it XMM-Newton} observation of this BeXRP. The paper is organised as follows: in Section \ref{sec:sec_obs_analyses}, we show observational information, the methodologies of our data reduction, and the results of spectral and temporal analyses. In Section \ref{sec:sec_discussion}, we present our discussions on the possible nature of the source and some relevant calculations, and in Section \ref{sec:conclusion}, we draw conclusions. 

\section{Observation and Analysis}
\label{sec:sec_obs_analyses}
\begin{figure}
    \centering
    \includegraphics[scale=0.3]{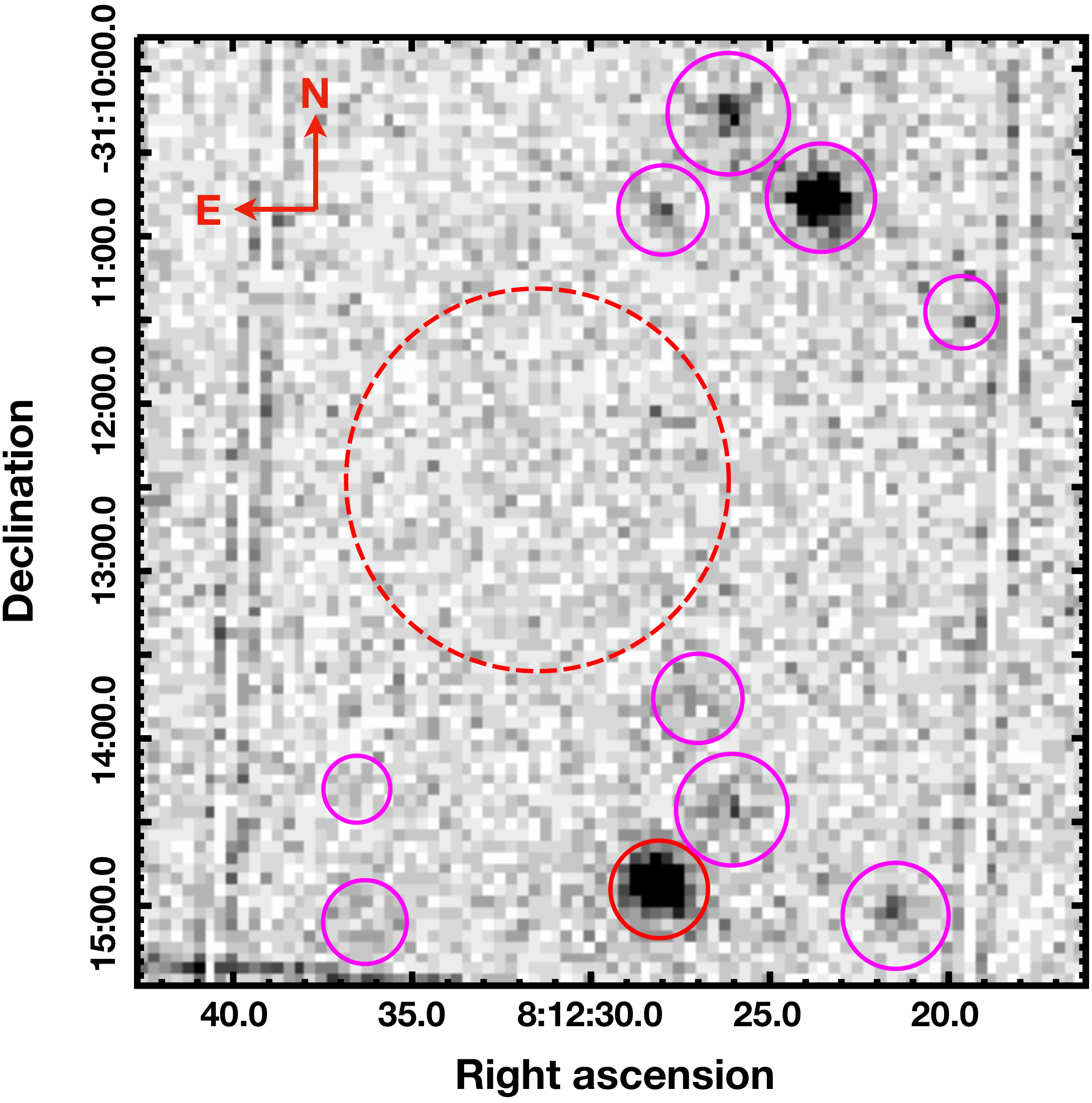}
    \caption{0.5-7 keV X-ray image from the EPIC-PN camera on {\it XMM-Newton}. The image is $340\arcsec \times 340\arcsec$ across. The source extraction region for J0812 is indicated with a solid red circle. The background region for extraction is indicated with a dashed red circle. Other nearby sources are indicated with solid magenta circles.}
    \label{fig:finding_chart}
\end{figure}

We use data from our \red{72-ks} {\it XMM-Newton} observation on 2018-10-09 using the European Photon Imaging Camera (EPIC; ObsID: 0822050101). Both PN and MOS detectors used full frame mode, and the source was covered by both detectors (see Fig. \ref{fig:finding_chart}). 
To avoid optical contamination, 
medium filters were applied to both PN and MOS cameras. We also made use of the 4.6 ks {\it Chandra} ACIS-S observation (ObsID: 14637; PI: Wijnands) taken on 2013-07-13.

For the {\it XMM-Newton} observation, we used the event files from the Processing Pipeline System (PPS) products, as derived from the Observation Data Files (ODF), for further reduction and analyses. \red{We cleaned the flaring particle background by first generating high energy (10-12 keV), single event (PATTERN=0) light curves for both PN and MOS cameras, using the {\tt evselect} task from the latest {\sc XMM Science Analysis Software} (SAS; version 17.0)\footnote{\url{https://www.cosmos.esa.int/web/xmm-newton/sas-download}}. Based on the light curves, we identified a period of low and steady background, with count rates $\leq 0.9~\mathrm{counts/s}$ for PN, $\leq 0.2~\mathrm{counts/s}$ for MOS1, and $\leq 0.4~\mathrm{counts/s}$ for MOS2. These thresholds were then applied to the {\tt tabgtigen} task to find good time intervals (GTIs), rendering effective exposures of $\approx$ 50 ks for PN, and of $\approx$ 63 ks for MOS1 and MOS2. The GTI files are then used to create filtered event files that are used to further generate spectra and time series.} The {\it Chandra} dataset was reprocessed with the {\tt chandra\_repro} task from {\sc CIAO} 4.11 (CALDB 4.8.2)\footnote{\url{http://cxc.cfa.harvard.edu/ciao/}} to align it with the up-to-date calibration. The reprocessed level-2 event file was used for further analyses.

To study the source's long term accretion history, we also obtained the {\it RXTE}/ASM light curve from the ASM Data Product page\footnote{\url{https://heasarc.gsfc.nasa.gov/docs/xte/asm_products.html}}. The light curve spans from 1996-01-05 to 2011-12-29.

\subsection{Spectral Analysis}
\label{sec:sec_spectral_fitting}
Spectra  extracted from the {\it XMM-Newton} and {\it Chandra} datasets were analysed with the {\sc HEASoft/Xspec} software. The PN and MOS spectra were rebinned to at least 20 counts per bin for simultaneous fits using $\chi^2$-statistics. \red{For each fit, we report the reduced $\chi^2$ ($\chi_\nu^2$ hereafter) together with the corresponding degrees of freedom (dof) as $\chi_\nu^2$ (dof).} Uncertainties and upper (or lower) limits of parameters are reported at the 90\% confidence level. The distance used in all analyses is the distance to the Be star (LS 992;  \citealt{motch1997}) obtained from the extended Gaia-DR2 distance catalogue ($d=6.76^{+1.20}_{-0.92}~\mathrm{kpc}$; \red{at the 68\% confidence level}), where distances are estimated from Gaia parallaxes by Bayesian analysis with a weak distance prior  \citep{bailer2018, Gaia18}. \red{Note that the distance estimate might be different using different types of measurement. For example, $d = 8.6 \pm 1.8~\mathrm{kpc}$ according to \cite{coleiro2013}. For all analyses of {\it XMM-Newton} spectra, we use the energy channels between $0.2$ and $10$ keV while channels between 0.5 and 10 keV are noticed for {\it Chandra} fits.} We accounted for interstellar absorption by convolving our models with the Tuebingen-Boulder ISM absorption model ({\tt tbabs} in {\sc Xspec}) using {\tt wilm} abundances (\citealt{wilms2000}). Since we have a moderately large number of spectral counts in the low energies, we tried fits with a free $n_\mathrm{H}$ (hydrogen column density), 
and fits with $n_\mathrm{H}$ fixed to the expected (based on HI) Galactic value ($\approx 0.48\times 10^{22}~\mathrm{cm^{-2}}$; see \citealt{kalberla2005}). 


We first tried a simple absorbed power-law ({\tt powerlaw} in {\sc XSPEC}, {\tt pow} hereafter) fit to the {\it XMM-Newton} spectra, finding a photon index ($\Gamma$) of $1.25\pm 0.10$. However, the fit exhibits strong residuals below $\sim$1 keV that lead to a $\chi_\nu^2$ of $1.68~(95)$, suggestive of an additional soft component. The $n_\mathrm{H}$ from this fit is a factor of $\sim 5$ below the Galactic value. However, if we force $n_\mathrm{H}$ to the Galactic value, the model gives a significantly worse fit with $\chi_\nu^2 = 2.66~(96)$ .

We proceeded by adding a blackbody component ({\tt pow + bbodyrad}) to account for the soft excess. The fit was significantly improved to a  $\chi_\nu^2$ of $1.003~(93)$, with a $kT=0.06\pm 0.01$ keV, a softer power-law index $\Gamma = 1.74^{+0.18}_{-0.17}$, and an enhanced absorption column density ($n_\mathrm{H}=1.25^{+0.24}_{-0.23}\times 10^{22}~\mathrm{cm^{-2}}$). The inferred radius of the blackbody emission region ($R_\mathrm{bb}$) reached $440.2^{+1223.6}_{-317.3}$ km -- too large to be consistent with the scale of a NS. The fit is significantly worse ($\chi_\nu^2= 1.31~(94)$) when $n_\mathrm{H}$ is fixed to the Galactic value; however this gives a much smaller emission region ($R_\mathrm{bb}= 10.3^{+5.6}_{-3.6}~\mathrm{km}$) and a slightly higher blackbody temperature ($kT_\mathrm{bb}= 0.10\pm 0.01~\mathrm{keV}$), which is consistent with the results from the 2013 {\it Chandra} observation. Similarly, a substitute for the soft component with a thermal bremsstrahlung model ({\tt pow + bremss}) also suggests an enhanced $n_\mathrm{H}=1.36^{+0.24}_{-0.23}\times 10^{22}~\mathrm{cm^{-2}}$, giving a good fit ($\chi_\nu^2= 1.02~(93)$), while fixing $n_\mathrm{H}$ to the Galactic value rendered a worse fit ($\chi_\nu^2 = 1.42~(94)$).

The soft component can also be modeled by a Gaussian emission line ({\tt pow + gauss}), with the line energy located at $0.63^{+0.06}_{-0.08}$ keV and a broad line width of $0.13^{+0.04}_{-0.03}$ keV when $n_\mathrm{H}$ is fixed. This model is 
acceptable 
either when $n_\mathrm{H}$ is free ($\chi_\nu^2= 0.99~(92)$) or fixed ($\chi_\nu^2 = 1.02~(93)$); however, no strong and broad emission features are expected in XRPs around this energy. 

We also tried fits with a magnetic NS atmosphere model ({\tt pow + nsmaxg}; see \citealt{ho2008, potekhin2014}), assuming a hydrogen atmosphere on a 1.4 $M_\odot$, 12 km NS, for different choices of magnetic field ($10^{10-13}$ G).  Initially, we assumed emission from the entire NS (normalisation$=(R_\mathrm{em}/R_\mathrm{NS})^2=1.0$, where $R_\mathrm{em}$ is the radius of the emission region). 
These fits were not superior to those using a blackbody. For example, when $B=10^{12}~\mathrm{G}$, the fit yielded $\chi_\nu^2= 1.61~(94)$ if $n_\mathrm{H}$ is freed ($n_\mathrm{H}= 0.26^{+0.05}_{-0.06}\times 10^{22}~\mathrm{cm^{-2}}$), and a worse $\chi_\nu^2 = 2.00~(95)$ if $n_\mathrm{H}$ is fixed. The former resulted in a surface temperature ($kT_\mathrm{s}$) of $0.07\pm0.01~\mathrm{keV}$, while the latter gives a slightly higher surface temperature at $0.082\pm 0.003~\mathrm{keV}$. We then tried fits with a free normalisation and found that they improved ($\chi_\nu= 1.13~(93)$ for $n_\mathrm{H}$-free v.s. $\chi_\nu= 1.45~(94)$ for $n_\mathrm{H}$-fixed); however, the inferred size of the emission region is too large to be plausible ($R_\mathrm{em}/R_\mathrm{NS} = 812^{+120}_{-401}$ for $n_\mathrm{H}$-free fit v.s. $R_\mathrm{em}/R_\mathrm{NS} = 6.2^{+2.9}_{-2.4}$ for $n_\mathrm{H}$-fixed fit). \red{We obtained better fits with $B=10^{11}~\mathrm{G}$. When the normalisation is fixed to unity, $\chi_\nu^2 = 1.46~(94)$ for free $n_\mathrm{H}$ ($=0.29^{+0.05}_{-0.05}\times 10^{22}~\mathrm{cm^{-2}}$) while $\chi_\nu^2 = 1.86~(95)$ for fixed $n_\mathrm{H}$. The $n_\mathrm{H}$-free fit gives a $kT=0.072^{+0.005}_{-0.007}~\mathrm{keV}$ while the $n_\mathrm{H}$-fixed fit results in a similar $kT=0.082^{+0.003}_{-0.003}~\mathrm{keV}$. With free normalisations, the fits are somewhat improved to $\chi_\nu^2=1.13~(93)$ when $n_\mathrm{H}$ is free and to $\chi_\nu^2 = 1.22~(94)$ when $n_\mathrm{H}$ is fixed. The problem with these fits, however, is still that the normalisations infer larger emission regions than the NS surface. The $n_\mathrm{H}$-free fit ($n_\mathrm{H} = 0.75^{+0.17}_{-0.15}\times 10^{22}~\mathrm{cm^{-2}}$) has quite a low $kT\leq 0.04~\mathrm{keV}$ that exceeds the allowed lower limit, so the corresponding inferred emission radius is much greater than the NS radius ($R_\mathrm{em}/R_\mathrm{NS}=36^{+40}_{-23}$); the $n_\mathrm{H}$-fixed fit gives a slightly higher $kT=0.04_{-0.01}^{+0.01}~\mathrm{keV}$ yet still yields an emission radius greater than the NS radius ($R_\mathrm{em}/R_\mathrm{NS}= 8^{+7}_{-3}$).}
We also tried to fix the normalisation to values smaller than unity, modelling surface hot spots, but did not find any significant 
improvement in the fits. 


We also substituted other physically motivated models for the power-law component. We first tried a Comptonisation model ({\tt comptt+bbodyrad}), assuming that soft photons from the blackbody component are up-scattered to form the hard component. We obtained a fair fit ($\chi_\nu^2 = 1.02~(91)$), with an optical depth ($\tau$) of $0.10^{+2.94}_{-0.08}$, but an unconstrained plasma temperature $kT_e \geq 36.80$ keV. The fit also suggests a higher $n_\mathrm{H}= 1.22^{+0.28}_{-0.22}\times 10^{22}~\mathrm{cm^{-2}}$, and gave a poorer fit ($\chi_\nu^2= 1.19~(92)$) when $n_\mathrm{H}$ was frozen at the Galactic value.

We found an equivalently good fit ($\chi_\nu^2= 1.01~(92)$ when $n_\mathrm{H}$ is free) using a power-law with an exponential high-energy cutoff ({\tt cutoffpl + bbodyrad}). When $n_\mathrm{H}$ was free, we found $\Gamma= 1.73^{+0.18}_{-0.74}$ but the cut-off energy was unconstrained ($E_\mathrm{cut}\geq 5.33$ keV).  The cutoff energy was better constrained to $2.97^{+2.93}_{-1.06}~\mathrm{keV}$ with $n_\mathrm{H}$ fixed to the Galactic value, but the fit itself became worse ($\chi_\nu^2= 1.19~(93)$).

The {\tt pow+bbodyrad}, {\tt comptt+bbodyrad}, {\tt cutoffpl+bbodyrad}, \red{and some of the {\tt pow+nsmaxg}} fits all statistically suggest an $n_\mathrm{H}$ above the Galactic value, but, with the high $n_\mathrm{H}$, infer a very high intrinsic luminosity of $\sim 10^{35}~\mathrm{erg~s^{-1}}$, which seems unlikely for a quiescent system. For example, the {\tt pow+bbodyrad} fit would give a blackbody component with $L_X\mathrm{\blue{(0.4-1~keV)}}\sim10^{35}$ erg/s, while the power-law component has a much smaller contribution ($L_X\mathrm{(\blue{1-10~keV})}\sim10^{33}$ erg/s). As the soft component must come from either reprocessed accretion energy (through either the NS surface, or an accretion disk), or other stored heat in the NS, it seems quite unlikely that the soft component could reach $\sim 10^{35}$ erg/s, while the hard component remains at $\sim 10^{33}$ erg/s. For this reason, we prefer the fits with fixed $n_\mathrm{H}$ on physical grounds.

Further investigation of the fixed-$n_\mathrm{H}$ fits indicates that the main reason for the poor fits is that the models are above the data at around 1 keV, which leaves apparent residuals that resemble an absorption feature, in both the PN and MOS spectra. 
We thus tried incorporating a Gaussian absorption line component ({\tt gabs}) into these models ({\tt pow+bbodyrad}, {\tt comptt+bbodyrad}, and {\tt cutoffpow+bbodyrad}) to compensate for the residuals. We found that the fits were 
significantly 
improved (e.g., the {\tt pow+bbodyrad} fit was improved from $\chi_\nu^2= 1.31~(94)$ to $\chi_\nu^2=1.09~(91)$; \blue{$\Delta\chi^2=23.95$}), so the absorption feature might be genuine. The resulting $kT_\mathrm{bb}$ from each model is slightly higher than in the original model without the {\tt gabs} component ($0.12\pm 0.01~\mathrm{keV}$ v.s. $0.11\pm 0.01~\mathrm{keV}$ for {\tt pow+bbodyrad} fit). The absorption line is regularly found between 0.99 and 1.02 keV.
As a point of comparison, we also added a {\tt gabs} component to the corresponding fits in which $n_\mathrm{H}$ was free, but found no clear improvement (\blue{$\chi_\nu^2= 1.00~(93)$ to $\chi_\nu^2=0.91~(90)$; $\Delta\chi^2=11.01$}). 

\red{The {\tt nsmaxg} models intrinsically include red-shifted cyclotron lines. For $B=10^{11}~\mathrm{G}$, the line is approximately at $0.94$ keV (assuming a 1.4-$M_\odot$, 12-km NS; this {\tt nsmaxg} model is 
henceforth
referred to as \# 1), which can partially compensate for the residual at 1 keV, so the {\tt pow + nsmaxg} fit is slightly better than the {\tt pow+bbodyrad} fit ($\chi_\nu^2 = 1.31~(94)$ vs. $\chi_\nu = 1.22~(94)$). The fit can be further improved by adjusting the line location; for example, one can use a larger radius to reduce the redshift and therefore elevate the line energy. We do find a better fit ($\chi_\nu^2 = 1.08~(94)$) when we increase $R_\mathrm{NS}$ from $12~\mathrm{km}$ to $14~\mathrm{km}$. A more physical approach might be using a model with the same mass and radius but a magnetic field slightly above $10^{11}~\mathrm{G}$ such that the intrinsic line locates exactly at $kT=1.01~\mathrm{keV}$ (as suggested by our {\tt gabs*(pow+bbodyrad)} fit, corresponding to a magnetic field of $1.07\times 10^{11}~\mathrm{G}$; this {\tt nsmaxg} model is henceforth referred to as \# 2). This also significantly improved the fit to $\chi_\nu^2 = 1.04~(94)$. However, either approach results in a normalisation greater than unity ($R_\mathrm{em}/R_\mathrm{NS}=6^{+6}_{-3}$ vs. $3.8^{+0.2}_{-0.3}$ for the former and latter cases, respectively). 
This model shows 
residuals at $\approx1.36$ keV, resembling 
a second 
absorption feature. We thus also tried convolving a {\tt gabs} component to the {\tt pow + nsmaxg} models. With the introduction of 3 more free parameters, the fits did not become any better but did yield more reasonable normalisations. For example, $\chi^2 = 0.95~(91)$ for model \# 2, with $R_\mathrm{em}/R_\mathrm{NS} = 1.3^{+2.9}_{-0.7}$. Introducing a second absorption line at a higher energy might be a result of variation of magnetic field due to complex distribution of the magnetic across the NS surface.}



We also fit the {\it Chandra} spectrum of RX J0812.4-3114 presented by \citet{tsygankov2017a} to a {\tt bbodyrad} model.  For this purpose, this low-count spectrum was binned to at least 1 count per bin, and we used C-statistics \citep{cash1979}. Because of the poorer calibration at low energies, all channels below 0.5 keV were ignored during the fit. We fixed $n_\mathrm{H}$ to the Galactic value, considering the discussion above, and the low-count statistics in this spectrum. The best-fitting model is a blackbody with a low temperature ($kT_\mathrm{bb}=0.13^{+0.03}_{-0.02}~\mathrm{keV}$) and an unconstrained blackbody radius ($R_\mathrm{bb}\leq 11.6~\mathrm{km}$). Adopting the Gaia-estimated distance of 6.76 kpc \citep{bailer2018, Gaia18}, we found an unabsorbed 0.5-10 keV luminosity of $5.5^{+5.2}_{-2.7}\times 10^{32}  (d/6.76~\mathrm{kpc})^2~\mathrm{erg~s^{-1}}$. \red{We also tried the {\tt nsmaxg} fits to the {\it Chandra} data, using both models \# 1 and \# 2, with free normalisations. We noted that there is no sign of absorption feature as in the {\it XMM-Newton} spectra. As a result, either model gives an equally fair fit (goodness$=47.4\%$ vs. goodness$=40.6\%$ for the \# 1 and \# 2, respectively). However, due to low counting statistics, we cannot get proper constraints on the normalisations.} To make sure that the absence of a hard component in the {\it Chandra} spectrum is not due to low counting statistics, we include a hard power-law component, with the power-law index ($\Gamma=1.32$) from the {\tt gabs*(pow+bbodyrad)} fit, and fit the normalisation parameter (with $kT_\mathrm{bb}$ and $R_\mathrm{bb}$ free) of the power-law. We found an upper limit for the power-law flux to be $\approx 2$ orders of magnitude lower than the flux of the blackbody component ($F_\mathrm{2-10}\lesssim 7.94\times 10^{-15}~\mathrm{erg~s^{-1}~cm^{-2}}$, or $L_\mathrm{2-10}\lesssim 4.34 \times 10^{31}\times (d/6.76~\mathrm{kpc})^2~\mathrm{erg~s^{-1}}$), suggesting that the {\it Chandra} spectrum is genuinely soft.

\begin{figure*}
    \centering
    \includegraphics[scale=0.5]{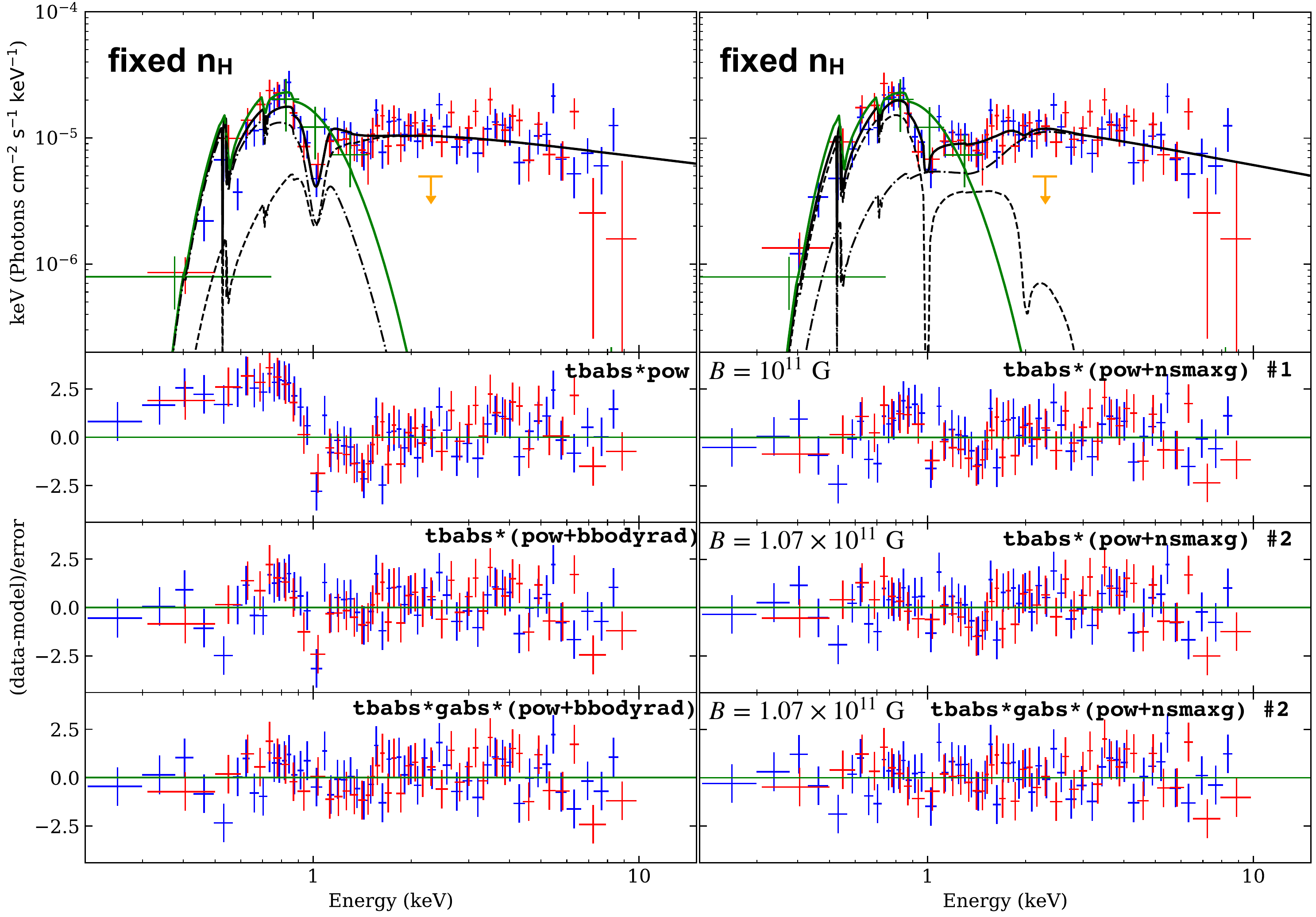}
    \caption{{\it XMM-Newton} spectra of PN (blue) and combined MOS (red) camera plotted together with the {\it Chandra} data (green). The orange bars with downward arrows on the two top panels mark the upper limit for the hard component in the {\it Chandra} spectrum as discussed in Sec. \ref{sec:sec_spectral_fitting}. The green solid lines in both of the top panels depict the best-fit {\tt tbabs * bbodyrad} model from the 2013 {\it Chandra} observation, consistent with the soft component alone from the {\it XMM-Newton} data. {\it Left panels:} {\it XMM-Newton} spectra with the soft excess modelled by {\tt bbodyrad} while $n_\mathrm{H}$ is fixed to the Galactic value. The top panel shows the best-fit model ({\tt tbabs * gabs * (pow + bbodyrad)}; solid black line), indicating the blackbody component (dashed-dotted) and a power-law component (dashed) separately. The lower panels show residuals resulting from attempted fits to different models. {\it Right panels:} {\it XMM-Newton} and {\it Chandra} spectra with the soft excess modelled by {\tt nsmaxg} at a fixed $n_\mathrm{H}$. The top panel shows the best-fitting model ({\tt tbabs * gabs* (pow+nsmaxg)}; solid black line), where the {\tt nsmaxg} and the power-law component are indicated with dashed and dashed-dotted black lines, respectively. The panels below show residuals from attempted fits to different models. }
    \label{fig:fig_xmmspec}
\end{figure*}

 \red{In summary, solely based on the fit quality, it seems that the {\it XMM-Newton} spectrum can be well-fit by either an absorbed soft model (which could be {\tt bbodyrad} or {\tt nsmaxg}) plus a hard component (which could be {\tt pow}, {\tt cutoffpl}, or {\tt comptt}) with increased absorption, or by the same model but with the $n_\mathrm{H}$ fixed to the Galactic value and modified by an absorption line (at $\approx 1.0$ keV for {\tt bbodyrad} and $\approx 1.3$ keV for {\tt nsmaxg}). However, considering the physical implications, the latter model is more favourable (see also Sec. \ref{sec:sec_discussion}). The inferred $kT_\mathrm{bb}$, $R_\mathrm{bb}$ (or $kT$ and $R_\mathrm{em}$ inferred from the {\tt nsmaxg} model), and thermal $L_X$ from the {\it XMM-Newton} data are 
 consistent with the results from the {\it Chandra} data.
 We summarise all relevant {\it XMM-Newton} spectral fitting parameters in Tab.~\ref{tab:tab_fit_results_xmm}, and 
 the {\it Chandra} parameters in 
 Tab.~\ref{tab:tab_fit_results_chandra}. In Fig.~\ref{fig:fig_xmmspec} we show the {\it XMM-Newton} spectra with the {\tt tbabs*gabs*(bbodyrad+pow)} and {\tt tbabs*gabs*(nsmaxg+pow)} models, with fixed $n_\mathrm{H}$, overplotted. For comparison, we also show the {\it Chandra} data and their best-fitting model ({\tt tbabs * bbodyrad}) and plot the upper limit of a possible hard component for the {\it Chandra} spectrum.}

\begin{table}
    \centering
    \resizebox{\columnwidth}{!}{%
    \renewcommand{\arraystretch}{1.29}
    \begin{tabular}{lccc}
    \hline
    Models   &   \multicolumn{3}{c}{Parameters} \\
             &    name & free $n_\mathrm{H}$ & fixed $n_\mathrm{H}$  \\
    \hline
    \multirow{5}{*}{{\tt tbabs*pow}}      & $n_\mathrm{H}~(10^{22}~\mathrm{cm^{-2}})$ & $0.11^{+0.04}_{-0.03}$ & $0.48^\dagger$ \\
                                          & $\Gamma$ & $1.25^{+0.10}_{-0.10}$ & $1.64^\ast$   \\
                                          & $F_X~(10^{-12}~\mathrm{erg~cm^{-2}~s^{-1}})$ & $0.15^{+0.01}_{-0.01}$ & $0.16^\ast$ \\
                                          & $L_X~(10^{33}~\mathrm{erg~s^{-1}})$ & $0.81^{+0.36}_{-0.23}$ & $0.89^{+0.34}_{-0.23}$ \\
                                          & $\chi_\nu^2$ (dof) & $1.684~(95)$ & $2.660~(96)$ \\  
    \hline
    \multirow{7}{*}{{\tt tbabs*(pow+bbodyrad)}}      & $n_\mathrm{H}~(10^{22}~\mathrm{cm^{-2}})$ & $1.25^{+0.24}_{-0.23}$ & $0.48^\dag$ \\
                                                     & $\Gamma$ & $1.74^{+0.18}_{-0.17}$ & $1.31^{+0.11}_{-0.11}$  \\
                                                     & $kT_\mathrm{bb}~(\mathrm{keV})$ & $0.06^{+0.01}_{-0.01}$ & $0.10^{+0.01}_{-0.01}$ \\
                                                     & $R_\mathrm{bb}~(\mathrm{km})$ & $440.2^{+1223.6}_{-317.3}$ & $10.3^{+5.6}_{-3.6}$\\
                                                     & $F_X~(10^{-12}~\mathrm{erg~cm^{-2}~s^{-1}})$ & $39.77^{+1.81}_{-1.81}$ & $0.36^{+0.02}_{-0.02}$ \\
                                                     & $L_X~(10^{33}~\mathrm{erg~s^{-1}})$ & $217.41^{+97.58}_{-62.27}$ & $1.95^{+0.88}_{-0.56}$ \\
                                                     & $\chi_\nu^2$ (dof) & $1.003~(93)$ & $1.308~(94)$ \\
    \hline
    \multirow{7}{*}{{\tt tbabs*(pow+gauss)}}         & $n_\mathrm{H}~(10^{22}~\mathrm{cm^{-2}})$ & $0.89^{+0.35}_{-0.34}$ & $0.48^\dag$ \\
                                                     & $\Gamma$ & $1.58^{+0.21}_{-0.21}$ & $1.37^{+0.10}_{-0.10}$ \\
                                                     & $E_\mathrm{gauss}~(\mathrm{keV})$ & $\lesssim 0.62$ & $0.63^{+0.06}_{-0.08}$ \\
                                                     & $\sigma_\mathrm{gauss}~(\mathrm{keV})$ & $0.19^{+0.08}_{-0.06}$ & $0.13^{+0.04}_{-0.03}$ \\
                                                     & $F_X~(10^{-12}~\mathrm{erg~cm^{-2}~s^{-1}})$ & $0.44^{+0.02}_{-0.02}$ & $0.22^{+0.01}_{-0.01}$ \\
                                                     & $L_X~(10^{33}~\mathrm{erg~s^{-1}})$ & $2.43^{+1.09}_{-0.70}$ & $1.23^{+0.55}_{-0.35}$ \\
                                                     &  $\chi_\nu^2$ (dof) & $0.990~(92)$ & $1.018~(93)$ \\
                                                     
    \hline
    \multirow{6}{*}{{\tt tbabs*(pow+bremss)}}        & $n_\mathrm{H}~(10^{22}~\mathrm{cm^{-2}})$ & $1.36^{+0.24}_{-0.23}$ & $0.48^\dag$ \\
                                                     & $\Gamma$ & $1.78^{+0.18}_{-0.18}$ & $1.30^{+0.11}_{-0.11}$ \\
                                                     & $kT_\mathrm{bremss}~(\mathrm{keV})$ & $0.08^{+0.01}_{-0.01}$ & $0.18^{+0.02}_{-0.02}$ \\
                                                     & $F_X~(10^{-12}~\mathrm{erg~cm^{-2}~s^{-1}})$ & $252.34^{+11.41}_{-11.41}$ & $0.50^{+0.02}_{-0.02}$ \\
                                                     & $L_X~(10^{33}~\mathrm{erg~s^{-1}})$ & $1379.35^{+618.21}_{-394.82}$ & $2.73^{+1.23}_{-0.78}$ \\
                                                     & $\chi_\nu^2$ (dof) & $1.018~(93)$ & $1.415~(94)$\\
    \hline
    \multirow{8}{*}{{\tt tbabs*(pow+nsmaxg)} \#1}    & $n_\mathrm{H}~(10^{22}~\mathrm{cm^{-2}})$ & $0.75^{+0.17}_{-0.15}$ & $0.48^\dag$  \\
                                                     & $\Gamma$ & $1.51^{+0.15}_{-0.16}$ & $1.36^{+0.11}_{-0.12}$\\
                                                     & $kT~(\mathrm{keV})$ & $\lesssim 0.04$ & $0.04^{+0.01}_{-0.01}$ \\
                                                     & $B~(\mathrm{G})$ & $10^{11}$ & $10^{11}$ \\
                                                     & $R_\mathrm{em}/R_\mathrm{NS}$ & $36.4^{+40.0}_{-23.0}$ & $7.8^{+7.4}_{-3.3}$ \\
                                                     & $F_X~(10^{-12}~\mathrm{erg~cm^{-2}~s^{-1}})$ & $1.40^{+0.07}_{-0.06}$ & $0.40^{+0.02}_{-0.02}$ \\
                                                     & $L_X~(10^{33}~\mathrm{erg~s^{-1}})$ & $7.67^{+3.45}_{-2.21}$ & $2.19^{+0.99}_{-0.63}$ \\
                                                     & $\chi_\nu^2$ (dof) & $1.130~(93)$ & $1.217~(94)$ \\
    \hline
    \multirow{8}{*}{{\tt tbabs*(pow+nsmaxg)} \#2}    & $n_\mathrm{H}~(10^{22}~\mathrm{cm^{-2}})$ & $0.52^{+0.09}_{-0.08}$ & $0.48^\dag$  \\
                                                     & $\Gamma$ & $1.27^{+0.11}_{-0.10}$ & $1.27^{+0.10}_{-0.10}$\\
                                                     & $kT~(\mathrm{keV})$ & $\lesssim 0.29$ & $\lesssim 0.29$ \\
                                                     & $B~(\mathrm{G})$ & $1.07\times 10^{11}$ & $1.07\times 10^{11}$ \\
                                                     & $R_\mathrm{em}/R_\mathrm{NS}$ & $4.2^{+0.8}_{-0.8}$ & $3.8^{+0.2}_{-0.3}$ \\
                                                     & $F_X~(10^{-12}~\mathrm{erg~cm^{-2}~s^{-1}})$ & $0.38^{+0.02}_{-0.02}$ & $0.34^{+0.02}_{-0.02}$ \\
                                                     & $L_X~(10^{33}~\mathrm{erg~s^{-1}})$ & $2.10^{+0.94}_{-0.60}$ & $1.88^{+0.84}_{-0.54}$  \\
                                                     & $\chi_\nu^2$ (dof) & $1.047~(93)$ & $1.043~(94)$ \\
    \hline
    \multirow{9}{*}{{\tt tbabs*(comptt+bbodyrad)}}   & $n_\mathrm{H}~(10^{22}~\mathrm{cm^{-2}})$ & $1.22^{+0.28}_{-0.22}$ & $0.48^\dag$  \\
                                                     & $kT_0~(\mathrm{keV})$ & $0.17^\ast$ & $0.43^{+0.15}_{-0.10}$ \\
                                                     & $kT_e~(\mathrm{keV})$ & $\gtrsim 36.8$ & $\lesssim 235.92$ \\
                                                     & $\tau$ & $0.10^{+2.94}_{-0.08}$ & $9.24^{+1.31}_{-7.89}$ \\
                                                     & $kT_\mathrm{bb}~(\mathrm{keV})$ & $0.06^{+0.01}_{-0.01}$ & $0.11^{+0.01}_{-0.01}$ \\
                                                     & $R_\mathrm{bb}~(\mathrm{km})$ & $440.6^{+1199.1}_{-332.0}$ & $8.3^{+4.3}_{-2.7}$ \\
                                                     & $F_X~(10^{-12}~\mathrm{erg~cm^{-2}~s^{-1}})$ & $39.94^{+1.80}_{-1.80}$ & $0.32^{+0.01}_{-0.01}$ \\
                                                     & $L_X~(10^{33}~\mathrm{erg~s^{-1}})$ & $218.36^{+97.87}_{-62.47}$ & $1.75^{+0.78}_{-0.50}$ \\
                                                     & $\chi_\nu^2$ (dof) & $1.023~(91)$ & $1.188~(92)$ \\
    \hline
    \multirow{8}{*}{{\tt tbabs*(cutoffpl+bbodyrad)}} & $n_\mathrm{H}~(10^{22}~\mathrm{cm^{-2}})$ & $1.24^{+0.20}_{-0.15}$ & $0.48^\dag$  \\
                                                     & $\Gamma$ & $1.73^{+0.18}_{-0.74}$ & $0.26^{+0.53}_{-0.58}$\\
                                                     & $E_\mathrm{cut}~(\mathrm{keV})$ & $\gtrsim 5.33$ & $2.97^{+2.93}_{-1.06}$ \\
                                                     & $kT_\mathrm{bb}~(\mathrm{keV})$ & $0.06^{+0.01}_{-0.01}$ & $0.11^{+0.01}_{-0.01}$ \\
                                                     & $R_\mathrm{bb}~(\mathrm{km})$ & $427.4^{+226.9}_{-339.7}$ & $8.7^{+4.7}_{-2.9}$ \\
                                                     & $F_X~(10^{-12}~\mathrm{erg~cm^{-2}~s^{-1}})$ & $38.13^{+1.72}_{-1.72}$ & $0.32^{+0.01}_{-0.01}$ \\
                                                     & $L_X~(10^{33}~\mathrm{erg~s^{-1}})$ & $208.44^{+93.42}_{-59.66}$ & $1.76^{+0.79}_{-0.50}$ \\
                                                     & $\chi_\nu^2$ (dof) & $1.014~(92)$ & $1.191~(93)$ \\
    \hline

    \end{tabular}
    }
    \caption{Best-fitting parameters to the {\it XMM-Newton} data. $F_X$ is unabsorbed flux over 0.2-10 keV, and $L_X$ is the corresponding luminosity. A $^\dag$ indicates parameters that were fixed during the fits, and a $^\ast$ marks parameters for which no valid constraints were found, thus should be taken with care. The strength in the {\tt gabs} model is related to the $\sigma$ and the optical depth ($\tau_\mathrm{gabs}$) of the line by $\mathtt{Strength}=\sqrt{2\pi}\sigma\tau_\mathrm{gabs}$. All {\tt nsmaxg} models assume a $1.4M_\odot$, $12~\mathrm{km}$ NS. All $kT_\mathrm{bb}$s, $R_\mathrm{bb}$s ($kT$s and $R_\mathrm{em}$s for {\tt nsmaxg} models), and $E_\mathrm{gabs}$s are redshifted quantities as observed by distant observers.}
    \label{tab:tab_fit_results_xmm}
\end{table}

\begin{table}
    \centering
    \resizebox{\columnwidth}{!}{%
    \renewcommand{\arraystretch}{1.3}
    \begin{tabular}{lccc}
    \multirow{10}{*}{{\tt tbabs*gabs*(pow+bbodyrad)}} & $n_\mathrm{H}~(10^{22}~\mathrm{cm^{-2}})$ & $1.01^{+0.25}_{-0.22}$ & $0.48^\dag$ \\
                                                          & $\Gamma$ & $1.60^{+0.19}_{-0.17}$ & $1.32^{+0.12}_{-0.12}$  \\
                                                          & $kT_\mathrm{bb}~(\mathrm{keV})$ & $0.08^{+0.01}_{-0.01}$ & $0.12^{+0.01}_{-0.01}$ \\
                                                          & $R_\mathrm{bb}~(\mathrm{km})$ & $88.7^{+249.5}_{-61.8}$ & $6.8^{+4.0}_{-2.2}$ \\
                                                          & $E_\mathrm{gabs}~(\mathrm{keV})$ & $1.00^{+0.03}_{-0.02}$ & $1.02^{+0.03}_{-0.04}$ \\
                                                          & $\sigma_\mathrm{gabs}~(\mathrm{keV})$ & $0.02^{+0.02}_{-0.01}$ & $0.05^{+0.04}_{-0.03}$ \\
                                                          & Strength & $0.22^{+0.65}_{-0.14}$ & $0.15^{+0.08}_{-0.06}$ \\
                                                          & $F_X~(10^{-12}~\mathrm{erg~cm^{-2}~s^{-1}})$  & $4.96^{+0.22}_{-0.22}$ & $0.32^{+0.01}_{-0.01}$ \\
                                                          & $L_X~(10^{33}~\mathrm{erg~s^{-1}})$  & $27.09^{+12.13}_{-7.75}$ & $1.78^{+0.80}_{-0.51}$ \\
                                                          & $\chi_\nu^2$ (dof) & 0.914 (90) & 1.085 (91) \\
     \hline
     \multirow{12}{*}{{\tt tbabs*gabs*(comptt+bbodyrad)}} & $n_\mathrm{H}~(10^{22}~\mathrm{cm^{-2}})$ & $0.86^{+0.32}_{-0.27}$ & $0.48^\dag$ \\
                                                     & $kT_0~(\mathrm{keV})$ & $0.13^\ast$ & $0.43^{+0.15}_{-0.10}$ \\
                                                     & $kT_e~(\mathrm{keV})$ & $\lesssim 80.98$ & $0.50^{+0.19}_{-0.12}$ \\
                                                     & $\tau$ & $9.28^{+0.10}_{-7.82}$ & $8.66^{+1.82}_{-7.11}$ \\
                                                     & $kT_\mathrm{bb}~(\mathrm{keV})$ & $0.09^{+0.03}_{-0.02}$ & $0.13^{+0.01}_{-0.01}$ \\
                                                     & $R_\mathrm{bb}~(\mathrm{km})$ & $36.8^{+138.5}_{-28.8}$ & $5.3^{+2.6}_{-1.8}$ \\
                                                     & $E_\mathrm{gabs}~(\mathrm{keV})$ & $0.99^{+0.02}_{-0.02}$ & $1.00^{+0.04}_{-0.03}$ \\
                                                     & $\sigma_\mathrm{gabs}~(\mathrm{keV})$ & $0.008^{+0.003}_{-0.003}$ & $0.05^{+0.04}_{-0.03}$ \\
                                                     & Strength & $\gtrsim 0.59$ & $0.16^{+0.08}_{-0.06}$ \\
                                                     & $F_X~(10^{-12}~\mathrm{erg~cm^{-2}~s^{-1}})$ & $1.77^{+0.08}_{-0.08}$ & $0.28^{+0.01}_{-0.01}$ \\
                                                     & $L_X~(10^{33}~\mathrm{erg~s^{-1}})$ & $9.68^{+4.34}_{-2.77}$ & $1.55^{+0.70}_{-0.44}$ \\ 
                                                     & $\chi_\nu^2$ (dof) & 0.910 (88) & 0.963 (89) \\
     \hline
    \multirow{11}{*}{{\tt tbabs*gabs*(cutoffpl+bbodyrad)}} & $n_\mathrm{H}~(10^{22}~\mathrm{cm^{-2}})$ & $0.86^{+0.30}_{-0.25}$ & $0.48^\dag$  \\
                                                     & $\Gamma$ & $0.94^{+0.73}_{-0.82}$ & $0.89^{+0.64}_{-0.73}$\\
                                                     & $E_\mathrm{cut}~(\mathrm{keV})$ & $\gtrsim 2.51$ & $2.68^{+2.79}_{-0.99}$ \\
                                                     & $kT_\mathrm{bb}~(\mathrm{keV})$ & $0.09^{+0.02}_{-0.01}$ & $0.13^{+0.01}_{-0.01}$ \\
                                                     & $R_\mathrm{bb}~(\mathrm{km})$ & $39.9^{+76.5}_{-18.2}$ & $5.6^{+3.4}_{-1.9}$ \\
                                                     & $E_\mathrm{gabs}~(\mathrm{keV})$ & $0.99^{+0.02}_{-0.02}$ & $1.01^{+0.04}_{-0.03}$ \\
                                                     & $\sigma_\mathrm{gabs}~(\mathrm{keV})$ & $0.010^{+0.003}_{-0.004}$ & $0.05^{+0.04}_{-0.03}$ \\
                                                     & Strength & $\gtrsim 0.59$ & $0.16^{+0.08}_{-0.06}$ \\
                                                     & $F_X~(10^{-12}~\mathrm{erg~cm^{-2}~s^{-1}})$ & $1.94^{+0.09}_{-0.09}$ & $0.29^{+0.01}_{-0.01}$ \\
                                                     & $L_X~(10^{33}~\mathrm{erg~s^{-1}})$ & $10.59^{+4.74}_{-3.03}$ & $1.56^{+0.70}_{-0.45}$ \\
                                                     & $\chi_\nu^2$ (dof) & 0.903 (89) & 0.96 (90) \\
    \hline
    \multirow{10}{*}{{\tt tbabs*gabs*(pow+nsmaxg)} \# 2}  & $n_\mathrm{H}~(10^{22}~\mathrm{cm^{-2}})$ & $0.58^{+0.16}_{-0.11}$ & $0.48^\dag$ \\
                                                          & $\Gamma$ & $1.58^{+0.54}_{-0.21}$ & $1.49^{+0.18}_{-0.19}$  \\
                                                          & $kT~(\mathrm{keV})$ & $\lesssim 0.09$ & $0.09^{+0.01}_{-0.02}$ \\
                                                          & $B~(\mathrm{G})$  & $1.07 \times 10^{11}$ & $1.07 \times 10^{11}$ \\
                                                          & $R_\mathrm{em}/R_\mathrm{NS}$ & $2.7^{+2.5}_{-1.4}$ & $1.3^{+2.9}_{-0.7}$ \\
                                                          & $E_\mathrm{gabs}~(\mathrm{keV})$ & $1.37^{+0.15}_{-0.84}$ & $1.37^{+0.11}_{-0.15}$ \\
                                                          & $\sigma_\mathrm{gabs}~(\mathrm{keV})$ & $0.39^{+0.25}_{-0.10}$ & $0.34^{+0.20}_{-0.07}$ \\
                                                          & Strength & $0.52^{+0.51}_{-0.25}$ & $0.65^{+0.37}_{-0.42}$ \\
                                                          & $F_X~(10^{-12}~\mathrm{erg~cm^{-2}~s^{-1}})$  & $0.42^{+0.02}_{-0.02}$ & $0.30^{+0.01}_{-0.01}$ \\
                                                          & $L_X~(10^{33}~\mathrm{erg~s^{-1}})$  & $2.28^{+1.02}_{-0.65}$ & $1.64^{+0.74}_{-0.47}$ \\
                                                          & $\chi_\nu^2$ (dof) & $0.941~(90)$ & $0.954~(91)$ \\
     \hline
    \end{tabular}
    }
    \contcaption{}
    \label{tab:tab_fit_results_xmm_continued}
\end{table}

\begin{table}
    \centering
    \begin{tabular}{lcc}
    \hline
    Model & Parameters & Values \\
    \hline
    \multirow{6}{*}{{\tt tbabs*bbodyrad}}   & $n_\mathrm{H}~(\mathrm{10^{22}~cm^{-2}})$ & $0.48^\dag$ \\
                                            & $kT_\mathrm{bb}~(\mathrm{keV})$ & $0.13^{+0.03}_{-0.02}$ \\
                                            & $R_\mathrm{bb}~(\mathrm{km})$ & $\lesssim 11.6$\\
                                            & $F_X~(\mathrm{10^{-12}~erg~s^{-1}~cm^{-2}})$ & $0.10^{+0.04}_{-0.03}$ \\
                                            & $L_X~(\mathrm{10^{33}~erg~s^{-1}})$ & $0.55^{+0.52}_{-0.27}$\\
                                            & Goodness & $17.0\%$ \\
    \hline
    \multirow{7}{*}{{\tt tbabs*nsmaxg} \#1} & $n_\mathrm{H}~(\mathrm{10^{22}~cm^{-2}})$ & $0.48^\dag$ \\
                                            & $kT~(\mathrm{keV})$ & $0.06^{+0.02}_{-0.01}$ \\
                                            & $B~(\mathrm{G})$ & $10^{11}$ \\
                                            & $R_\mathrm{em}/R_\mathrm{NS}$ & $\lesssim 10.4$\\
                                            & $F_X~(\mathrm{10^{-12}~erg~s^{-1}~cm^{-2}})$ & $0.12^{+0.05}_{-0.04}$ \\
                                            & $L_X~(\mathrm{10^{33}~erg~s^{-1}})$ & $0.67^{+0.64}_{-0.33}$\\
                                            & Goodness & $47.4\%$ \\
    \hline
    \multirow{7}{*}{{\tt tbabs*nsmaxg} \#2} & $n_\mathrm{H}~(\mathrm{10^{22}~cm^{-2}})$ & $0.48^\dag$ \\
                                            & $kT~(\mathrm{keV})$ & $\lesssim 0.10$ \\
                                            & $B~(\mathrm{G})$ & $1.07\times10^{11}$ \\
                                            & $R_\mathrm{em}/R_\mathrm{NS}$ & $5.6^\ast$\\
                                            & $F_X~(\mathrm{10^{-12}~erg~s^{-1}~cm^{-2}})$ & $0.10^{+0.04}_{-0.03}$ \\
                                            & $L_X~(\mathrm{10^{33}~erg~s^{-1}})$ & $0.54^{+0.53}_{-0.27}$\\
                                            & Goodness & $40.6\%$ \\
    \hline
    \end{tabular}
    \caption{Best-fitting parameters to the 2013 {\it Chandra} spectrum. The notations for subscripts and superscripts are the same as in Tab. \ref{tab:tab_fit_results_xmm}.  $F_X$ is the unabsorbed flux over 0.5-10 keV, and $L_X$ is the corresponding unabsorbed luminosity.} 
    \label{tab:tab_fit_results_chandra}
\end{table}


\subsection{Temporal Analyses}
The PN camera has sufficiently high timing resolution (73.4 ms in full-frame mode) to search for pulsations in this system. For that purpose, we first applied barycentric correction for the arrival times of photons in the flare-cleaned event lists, and then extracted PN light curves with SAS {\tt evtselect} task over a soft band that primarily covers the soft excess (0.4-1.0 keV), a band covering the hard spectral component (1-10 keV), and a broad band (0.4-10 keV). These time series were then rebinned to 1 s time bins, which are short enough to search for the expected pulse period ($\approx 31.88$ s; see \citealt{corbet2000}).

\begin{figure*}
	\includegraphics[scale=0.5]{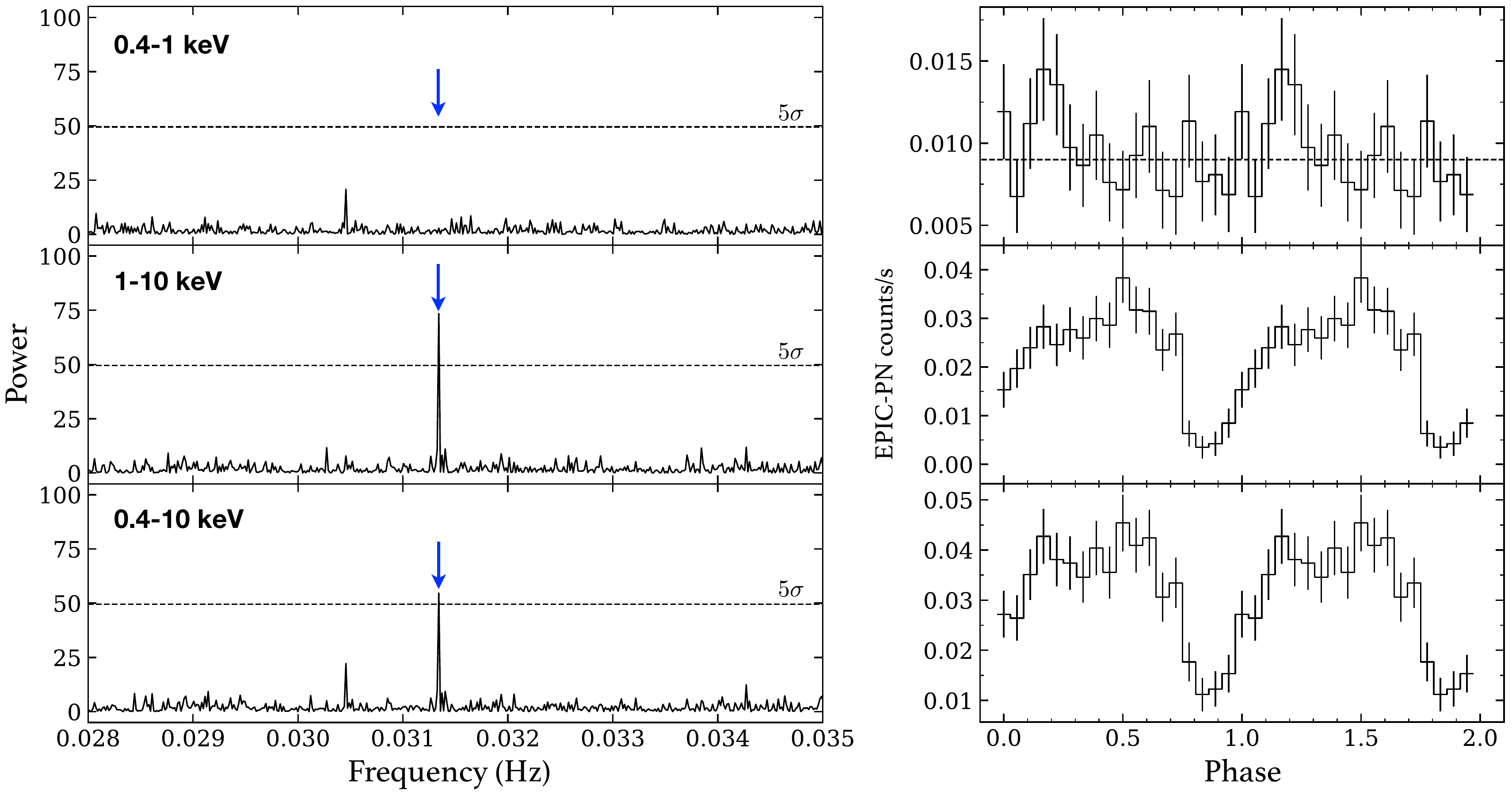}
     \caption{{\it Left}: power spectra of PN time series. The frequency that corresponds to the best period ($\approx 31.908$ s) found by {\tt efsearch} is indicated with an arrow in each panel; no signal is present in the soft band (top). The dashed line in each panel depicts the $5\sigma$ level. {\it Right}: Corresponding light curves folded at the best period.  In the top panel (also 0.4-1.0 keV), for the soft excess, the dashed horizontal line depicts the best-fitting constant to the light curve.}
     \label{fig:powspec}
\end{figure*}

\begin{figure}
    \centering
    \includegraphics[scale=0.44]{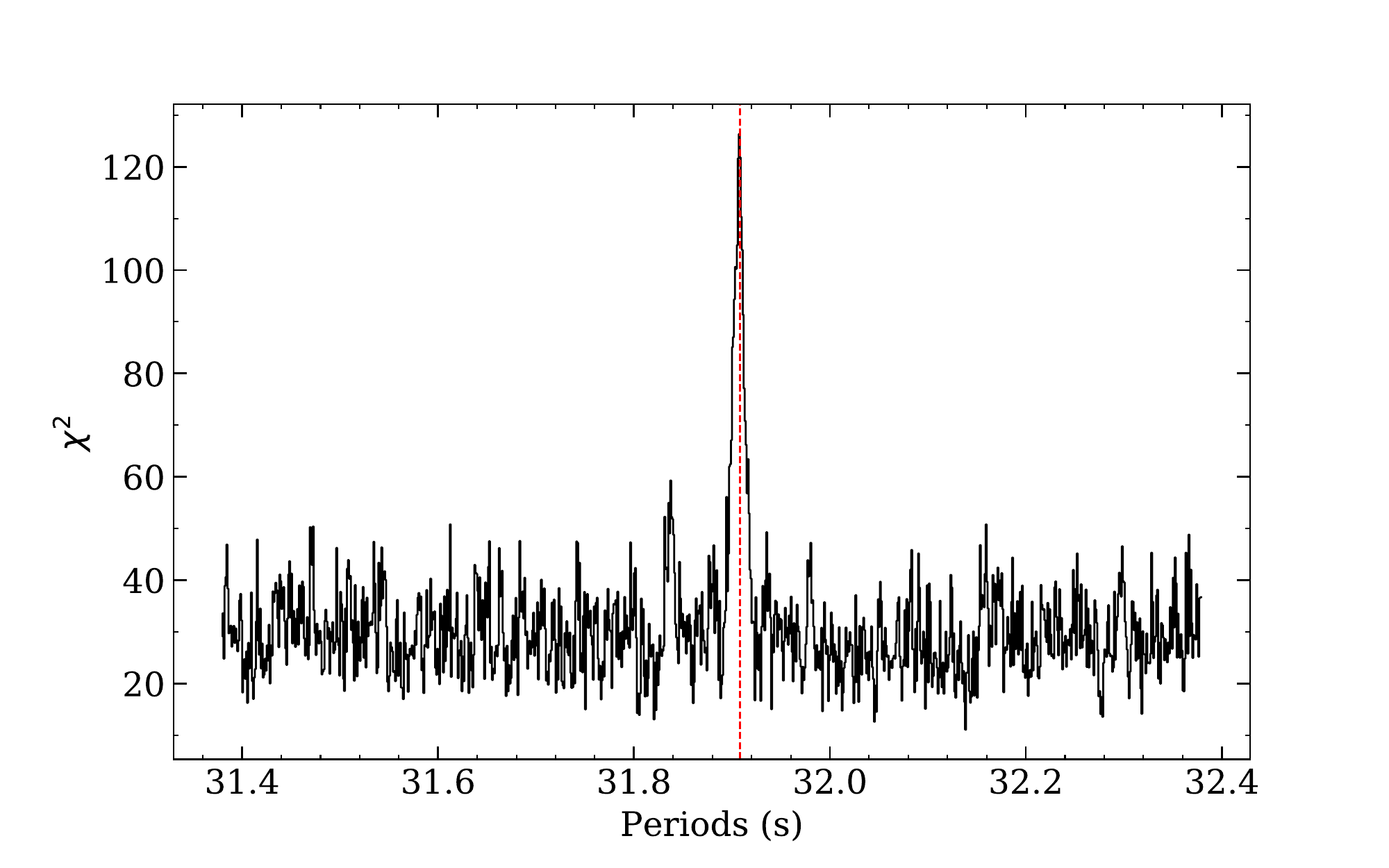}
    \caption{$\chi^2$s of fitting folded EPIC-PN light curves to constants calculated by {\tt efsearch} vs. the periods used for folding. The best period found by {\tt efsearch} is indicated with a dashed red line.}
    \label{fig:p_search_pn}
\end{figure}

Timing analyses were performed with tasks from the {\sc HEASARC/Xronos} software package\footnote{\url{https://heasarc.gsfc.nasa.gov/xanadu/xronos/xronos.html}}. We searched for pulsations by running the {\tt powspec} task on the rebinned PN light curve in each band \blue{using a total of 32768 frequency bins between $1.52\times10^{-5}$ and $0.5~\mathrm{Hz}$}. A clear periodicity was revealed at $\approx 0.031~\mathrm{Hz}$ in the hard band power spectrum, represented by a prominent peak (\blue{power$\approx 72.53$}; see left panels of Fig. \ref{fig:powspec}). \blue{The noise in a power spectrum is expected to follow a $\chi^2$ distribution with $2$ degrees of freedom \citep{Leahy83}, from which we derived a 5$\sigma$ significance level given the number of frequency bins in our analysis}. To search for the exact period, we used the {\tt efsearch} tool, which rebins and folds the light curve over a range of period and searches for best period by finding the maximum $\chi^2$ from fitting a constant to the folded light curves (see Fig. \ref{fig:p_search_pn}). We then folded the light curves at the best period with the {\tt efold} task to create pulse profiles (see right panels of Fig. \ref{fig:powspec}).

We found a best pulse period in the hard-band light curve at $31.908\pm 0.009$ s (upper and lower bounds correspond to periods with $\chi^2$ values at half of the maximum). Compared to the pulse period ($P_{1999}\approx 31.8856 \pm 0.0001~\mathrm{s}$) found by \cite{corbet2000}, this indicates a spin-down ($\dot P$) of $3.63\times 10^{-11}~\mathrm{s~s^{-1}}$. This spin-down is rather slow, but not particularly unusual in BeXRPs. For example, SAX J0635+0533 was observed to have a long-term spin down of $>3.8\times 10^{-13}$ since its discovery \citep{palombara2017}. However, the difference in spin period could have been affected by the orbital Doppler effect. To estimate the maximal magnitude of the orbital Doppler effect, we assume the system is edge-on, and use the primary mass of $17~\mathrm{M_\odot}$ from \cite{reig2001}. The resulting Lorentz factor due to orbital motion ($\approx 4\times 10^{-4}$) introduces an uncertainty of $0.013~\mathrm{s}$ in the spin period. This is comparable to the spin difference we have measured, so, without further knowledge of the orbital ephemeris and inclination, the spin-down reported here should be interpreted with caution.

We found no clear signs of pulsations in the \blue{folded} light curve of the soft excess, to which we fitted a constant and found a $\chi_\nu^2=0.76$. However, clear sharp dips are present in the folded hard and broad band light curves (see Fig. \ref{fig:powspec}), which were previously suggested by \cite{galloway2001} to be partial eclipses of the emitting region by the channeled accretion column.
 
 To quantify the light curve modulation, we calculated the pulsed fraction, which is defined as 

\begin{equation}
    \mathrm{PF} = \dfrac{C_\mathrm{max} - C_\mathrm{min}}{C_\mathrm{max} + C_\mathrm{min}},
    \label{eq1}
\end{equation}
where $C_\mathrm{max}$ and $C_\mathrm{min}$ are the maximum and the minimum count rates, respectively. We found $\mathrm{PF}= 0.84 \pm 0.10$ for the hard band, whereas the pulse fraction for the soft band is more uncertain but 
could be very low 
($\mathrm{PF} = 0.37 \pm 0.18$). Non-detection of pulsations in the soft excess could be a result of the relatively low counting statistics ($\approx 323~\mathrm{counts}$), or due to the fact that the pulsed fraction is genuinely too low. To test this, we generated a series of simulated light curves that are modulated by sinusoidal functions at the observed pulsed period ($31.908$ s) but with different pulse fractions (up to 0.99). Using the observed counts in the soft excess, we generate 100 lightcurves for each pulsed fraction. For each pulsed fraction, pulsations are considered to be detectable if more than 90\% of the realisations result in powers at the expected pulsed period that are 3 $\sigma$ above the corresponding noise levels. With the given counts in the soft excess, we found that the pulsed fraction has to be $\gtrsim 31\%$ for a pulsed signal to be detected. In other words, if the pulsation is not detected, the pulsed fraction is then at most $31\%$. 

\begin{figure}
    \centering
    \includegraphics[scale=0.52]{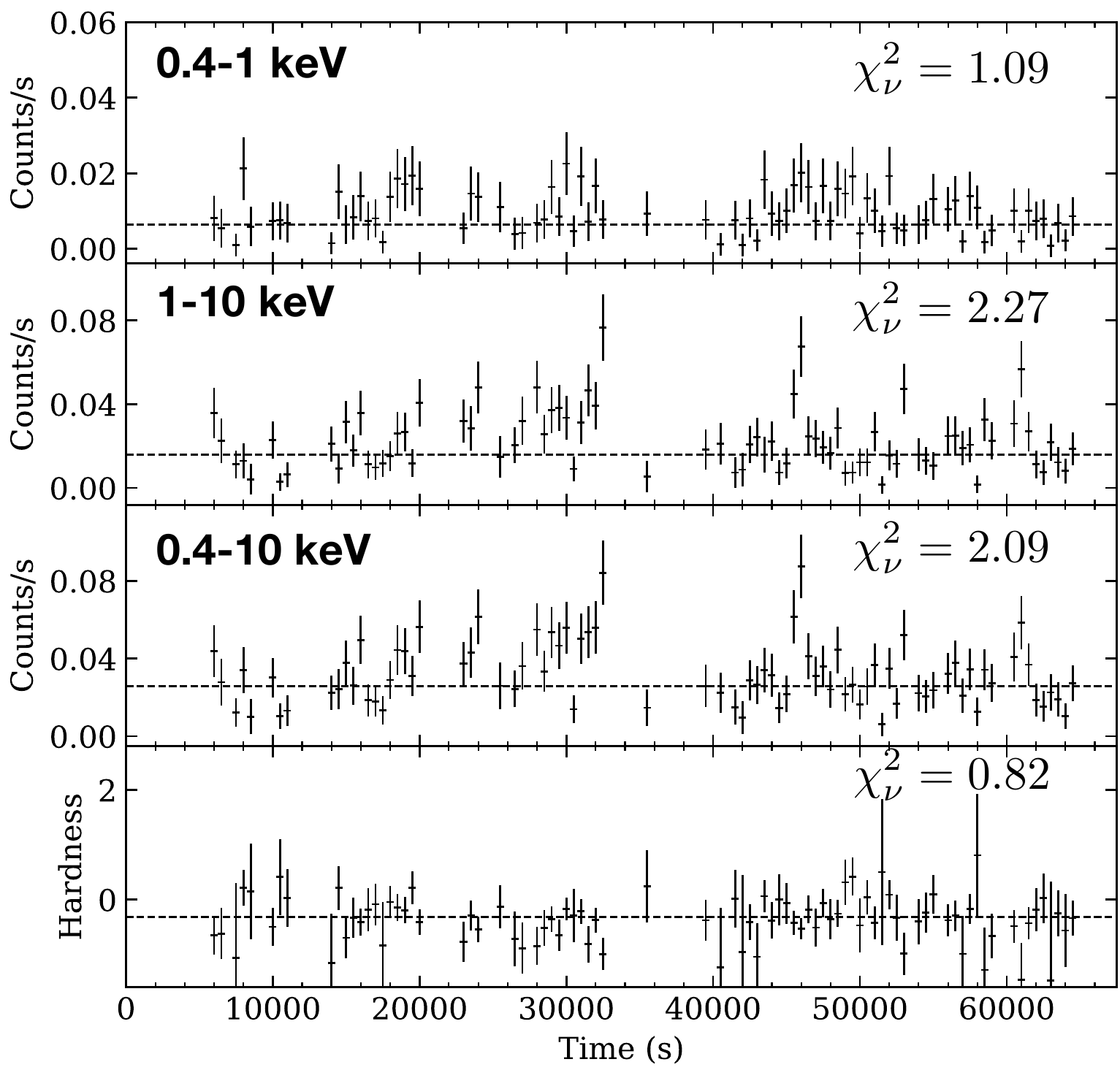}
    \caption{Time series rebinned to 500 s time bins. Light curves in soft (0.4-1.0 keV), hard (1-10 keV) and broad (0.4-10 keV) are shown in the first, second and third panels, respectively. The bottom panel presents the time series of the hardness ratio defined by eq.(\ref{eq1}). Errors in the hardness ratio are propagated from errors in the soft and hard count rates. The dashed line in each panel marks the best-fitting constant.}
    \label{fig:long_term_lc}
\end{figure}

To check for long-term variability, we rebinned the light curves to 500 s time bins, using the same set of soft (0.4-1 keV), hard (1-10 keV), and broad (0.4-10 keV) bands while defining a hardness ratio with

\begin{equation}
    \text{Hardness} = \log_{10}\left(\frac{C_{0.4-1}}{C_{1-10}}\right).
    \label{eq2}
\end{equation}
We then fitted these rebinned light curves to constants and used the resulting reduced $\chi^2$s as a measure of variability. Fig. \ref{fig:long_term_lc} shows the rebinned light curves and the corresponding time series of hardness ratios.

\begin{figure}
    \centering
    \includegraphics[scale=0.8]{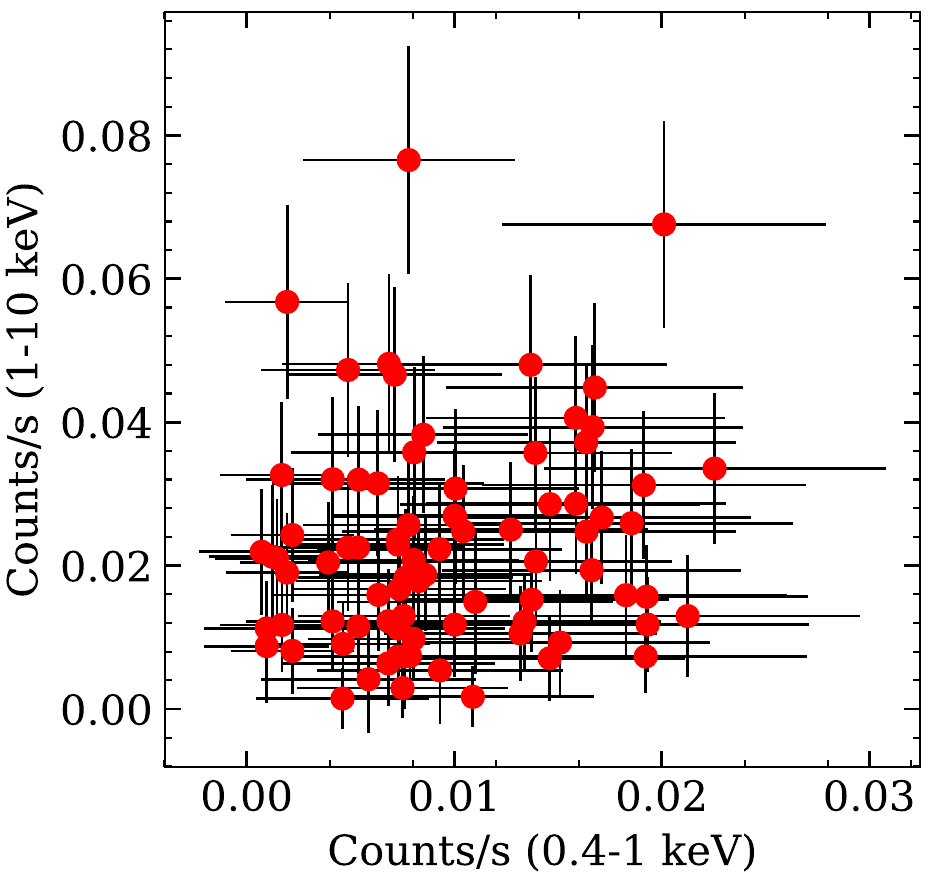}
    \caption{Hard count rates v.s. the corresponding soft count rates. No clear correlation was found between the hard count rate and the soft count rate, so the soft emission might have a completely different origin from the hard emission.}
    \label{fig:fig_hard_vs_soft}
\end{figure}

The result implies strong variability in the hard band light curve ($\chi_\nu^2\approx 2.27$), while the soft excess shows no sign of variability ($\chi_\nu^2\approx 1.09$). Due to the large error bars, variability in the hardness ratio is hard to determine solely based on the $\chi_\nu^2$ ($\approx 0.82$); however, the best-fitting hardness ratio ($\approx -0.32^{+0.04}_{-0.04}$) suggests that the soft excess contributes less than 50\% of the total observed flux ($F_{0.4-1}/F_{0.4-10}\approx 32.3\%$). To further explore a possible correlation between the soft and hard counts, we calculated a correlation coefficient defined as

\begin{equation}
    \rho = \dfrac{\mathrm{Cov}(C_\mathrm{soft}, C_\mathrm{hard})}{\sigma_\mathrm{soft}\sigma_\mathrm{hard}},
    \label{eq3}
\end{equation}
where $\mathrm{Cov}(C_\mathrm{soft}, C_\mathrm{hard})$ is the covariance between the soft and hard count rates, while $\sigma_\mathrm{soft}$ and $\sigma_\mathrm{hard}$ are standard deviations in soft and hard rates, respectively. $\left|\rho\right| = 1$ corresponds to linear correlation, and $\rho = 0$ corresponds to non-correlation. We found $\rho = 0.15 \pm 0.14$ (the uncertainty is propagated from the data), suggesting very weak or no linear correlation between the soft and hard count rates. The soft counts might therefore have a completely distinct origin from the hard counts. A plot of soft count rates against their corresponding hard count rates can be found in Fig. \ref{fig:fig_hard_vs_soft}.

\begin{figure}
    \centering
    \includegraphics[scale=0.44]{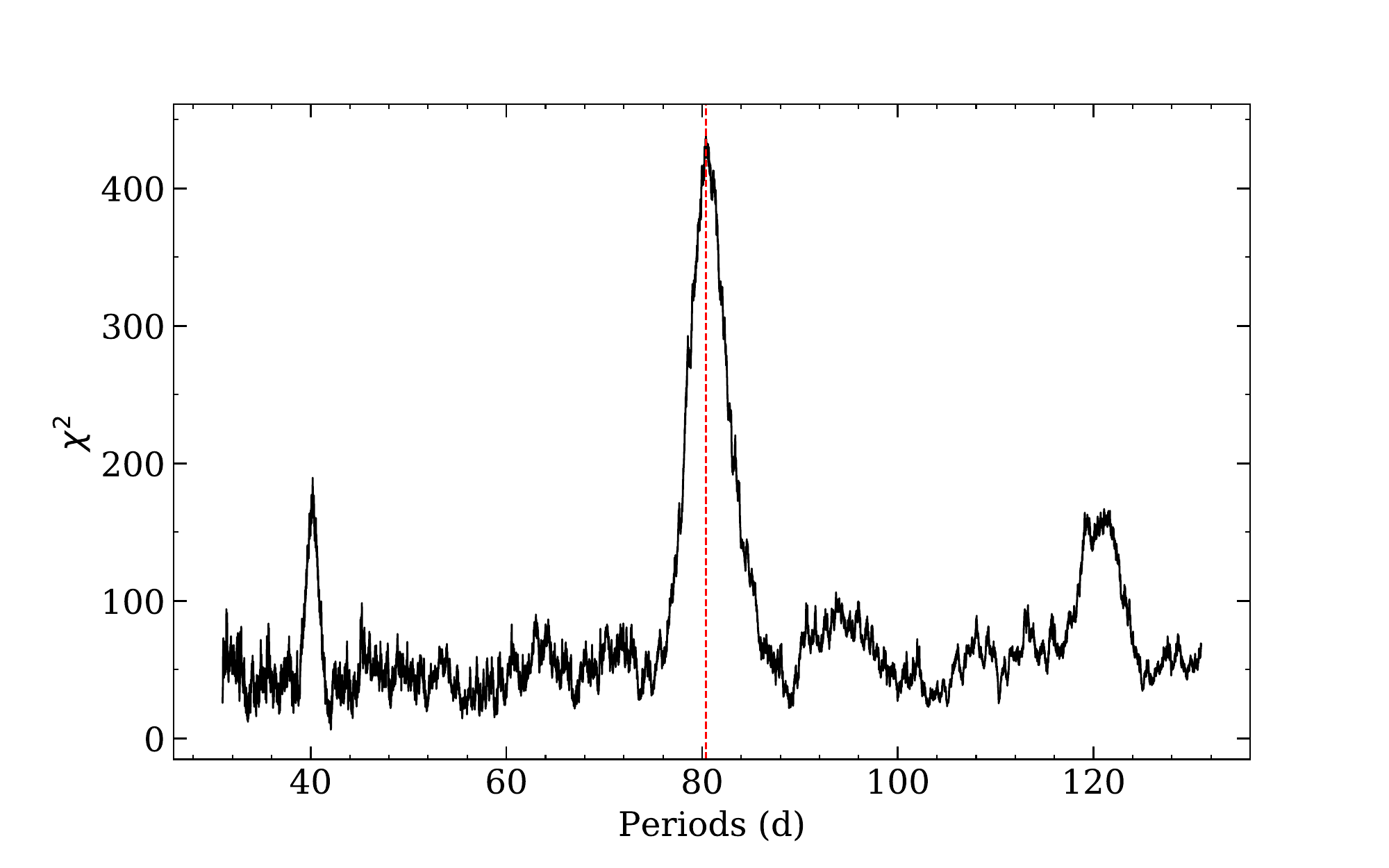}
    \caption{$\chi^2$s of fitting folded {\it RXTE/ASM} light curves (during the active period) to constants calculated by {\tt efsearch} vs. the periods used for folding. The best period found by {\tt efsearch} is indicated with a dashed red line.}
    \label{fig:p_search_asm}
\end{figure}

\begin{figure*}
    \centering
    \includegraphics[scale=0.5]{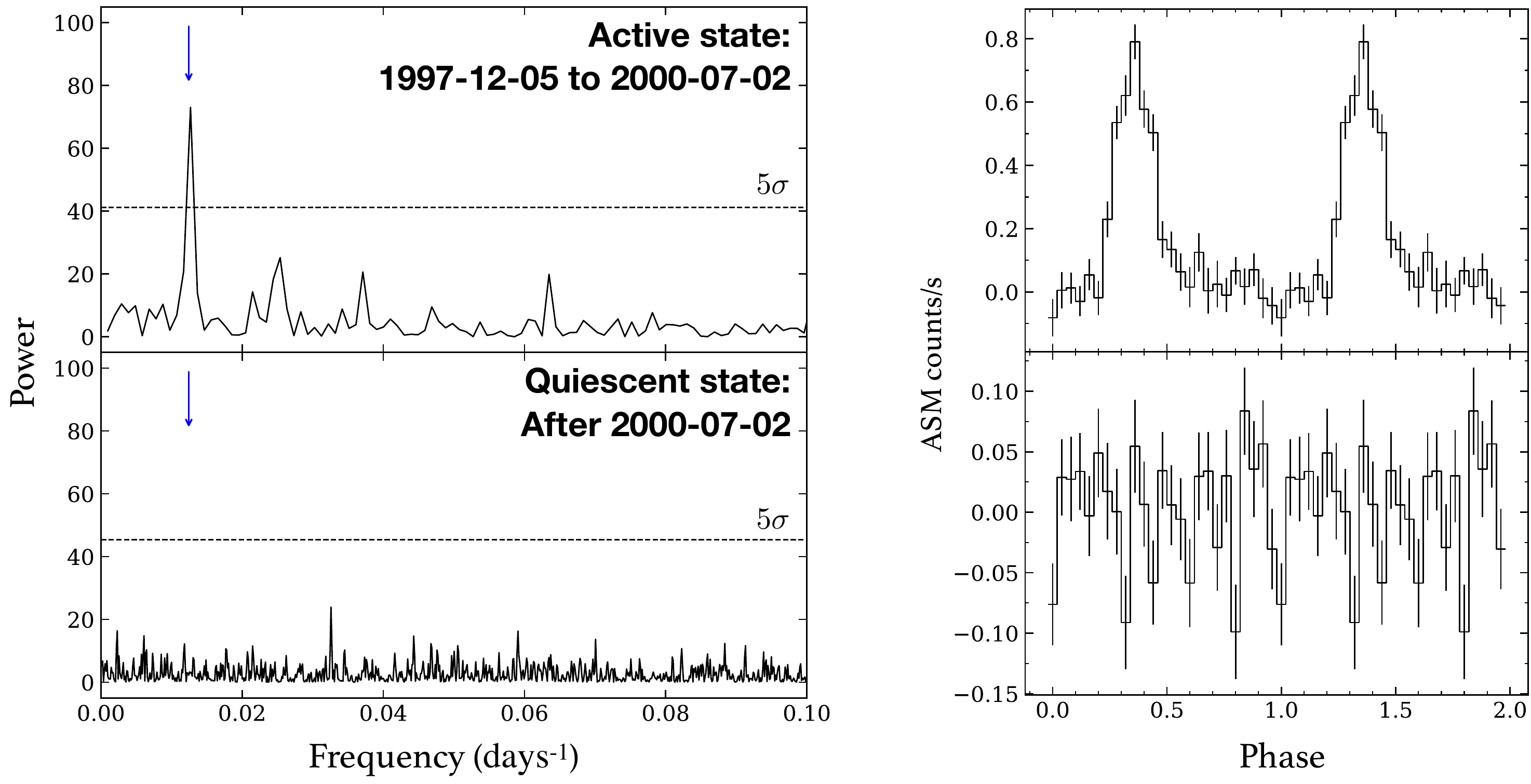}
    \caption{{\it Left:} ASM power spectra calculated with data from the active epoch (top) and data from the quiescent epoch (bottom). The dashed line in each panel depicts the $5\sigma$ level. The best period ($\approx 80.393~\mathrm{days}$) is indicated with a blue arrow. {\it Right:} corresponding light curves folded at the best period found by {\tt efsearch}.}
    \label{fig:asm_pow_lc}
\end{figure*}

To see if the source returned to quiescence after the active state, we also compared the {\tt powspec} and {\tt efsearch} results on the ASM light curve during the active state (between 1997-12-05 and 2000-07-02) with those on the light curve after the active state. We found a clear periodicity only in the former case. The best period was found to be $P_\mathrm{orb}=80.39^{+3.00}_{-2.18}$ days for the active epoch with a maximum power of $73.06$ (uncertainties in the period are estimated using periods with $\chi^2$ values that are half of the maximum; see Fig. \ref{fig:p_search_asm}). This is consistent with the $\sim 81.3$-day orbital period found by \cite{corbet2000}. Because the ASM light curves are background-subtracted, some phases contain negative count rates. We approximated a backgrond level by fitting the quiescent light curve to a constant. This gives a best-fit value at $-0.11~\mathrm{counts/s}$ with $\chi_\nu^2= 1.77$, indicating variability possibly due to some minor source activity. We then applied this background level to the light curves to shift them to unsubtracted levels. The power spectra of the active and quiescent epochs are shown in the left panels of Fig. \ref{fig:asm_pow_lc} (\blue{the 5$\sigma$ level is calculated using the same method as for the PN power spectra}), while the corresponding light curves folded at the best period are shown in the right panels.

\section{Discussion}
\label{sec:sec_discussion}
\subsection{Soft Excess}

The 2013 {\it Chandra} observation revealed a very soft ($F_{2-10}/F_{0.5-10}\lesssim 4.1\times 10^{-4}$) spectrum which makes RX J0812.4-3114 the coolest ($kT_\text{bb}\sim 0.1~\mathrm{keV}$) among all known quiescent BeXRPs \citep{tsygankov2017a} that share similar spectral shapes. Although not well constrained due to low counting statistics, the fit to the {\it Chandra} data does suggest a blackbody spectrum with a large emitting region ($R_\mathrm{bb}\sim 10~\mathrm{km}$), which has never been observed in any other quiescent BeXRP.

However, soft excesses have been observed in some luminous ($L_X \gtrsim 10^{37}~\blue{\mathrm{erg~s^{-1}}}$) XRPs. Particularly in high-inclination systems, soft X-rays are thought to originate from reprocessing of hard X-rays by the optically-thick inner disc region, which leads to a larger effective $R_\mathrm{bb}$ (e.g., \citealt{endo2000}). \cite{hickox2004} have estimated that the corresponding blackbody temperature ($T_\mathrm{bb}$) is related to $L_X$ by $T_\mathrm{bb}\propto L_X^{11/28}$, so in faint XRPs ($L_X\lesssim 10^{36}~\blue{\mathrm{erg~s^{-1}}}$), given that $L_X$s are low, the majority of the reprocessed hard X-rays would instead shift into the EUV regime. Reprocessing of hard X-rays is therefore not likely to be the mechanism at work for the soft excess in J0812. It might seem to be possible that the intrinsic absorption suggested by the $n_\mathrm{H}$-free fits might arise from an obscured inner disc region. The high unabsorbed luminosity mentioned in Sec. \ref{sec:sec_spectral_fitting} for these fits might therefore favour this scenario. However, because the soft excess is powered by the hard X-rays, the scenario is valid only when the hard component is brighter or at least comparable to the soft excess (e.g., \citealt{burderi2000} found a soft excess in Cen X-3 that takes $\approx 58\%$ of the total unabsorbed flux). Moreover, \cite{endo2000} showed that the cooling timescale of the irradiated inner disc should be only a fraction of a second, so the soft excess should also be pulsed in accordance with the hard component. Therefore, the above discussion strongly disfavours the $n_\mathrm{H}$-free fits with their enhanced absorption.

The companion Be star may partially contribute to the soft X-rays. According to \cite{naze2011}, we can roughly estimate the expected $L_X$ from the companion star adopting the reported B0.2IVe spectral type  (Sec. \ref{sec:sec_intro}), which suggests that the companion's X-ray luminosity should lie in the range of $10^{31-31.5}~\mathrm{erg~s^{-1}}$. This is $\sim 2$ orders of magnituides lower than the observed luminosity, so unlikely to be a significant contributor to the soft component. 

Soft X-ray emission in quiescence has typically been ascribed to thermal emission from the NS, either from small regions of higher temperature -- hot spots -- or, although not previously detected in a quiescent BeXRP, from the entire NS surface. Hot spots can either be formed externally as channeled accretion columns heat up the polar caps (e.g. pulsed soft excess was noted in the BeXRP RX J1037.5-5647 by \citealt{palombara2009} with $R_\mathrm{bb}\sim 128~\mathrm{m}$), 
or intrinsically as heat from the core is channeled toward parts of the surface by strong internal magnetic fields \citep{greenstein1983, potekhin2001, geppert2004}. The large inferred $R_\mathrm{bb}$ from our analyses therefore indicates a rather large hot spot. The former is then not likely the mechanism at work since the spot size in this source is actually predicted to be $\sim 0.1~\mathrm{km}$ (following $R_\mathrm{pc}=(2\pi R_\mathrm{NS}/(c P))^{1/2} R_\mathrm{NS}$, e.g. \citealt{LyneGrahamSmith2006}, \citealt{forestell2014}).
Large hot spots of radius several km have indeed been detected in some NSs (e.g., \citealt{gotthelf2009}). However, absence of pulsations in the soft excess weakens the hot spot scenario, although it is still possible to have a relatively small pulsed fraction with a particular observer geometry. The large inferred $R_\mathrm{bb}\sim 10~\mathrm{km}$ in our source suggests that the soft X-ray photons might have primarily originated from the whole NS surface. Even our fits with the smallest $R_\mathrm{bb}$ (e.g., $R_\mathrm{bb}\approx 5.3^{+2.6}_{-1.8}~\mathrm{km}$ in the {\tt bbodyrad + comptt} fit) are much larger than predicted hot spot sizes, indicating that the observed spectrum might be comprised of a hot spot plus emission from the overall surface (see e.g. \citealt{elshamouty2016}.) 


Heat thermally radiated during quiescent states is thought to be principally deposited during the previous accretion episodes, especially the bright outbursts. After outbursts, NSs cool via thermally radiating away the heat from the surface, and/or through either slow (e.g., modified Urca) or fast (e.g., direct Urca) neutrino emission processes in the core (\citealt{potekhin2015}, and references therein).
Heat is generated both by pycnonuclear reactions in the deep crustal regions (``deep crustal heating"; see \citealt{brown1998}), which leaks out over long ($\sim10^5$ year) timescales, and by several processes in the outer crust, which leak out of the NS on shorter (months to decades) timescales \citep{Rutledge02c,Shternin07,Brown09,deibel2015}.
It is usually assumed that the NS crust and core return back to thermal equilibrium several years after the outbursts (see the review by \citealt{wijnands2017}). The heating rate ($H$) in this case is then simply related to the quiescent luminosity ($L_q$) and the neutrino cooling rate ($L_\nu$) by

\begin{equation}
    H = L_q + L_\nu.
    \label{eq4}
\end{equation}
Here, $H$ depends on the average mass accretion rate of the system:

\begin{equation}
    H = \dfrac{\langle \dot M\rangle}{m_u} Q_\mathrm{nuc},
    \label{eq5}
\end{equation}
where $Q_\mathrm{nuc}$ is the amount of energy generated by pycnonuclear reactions per accreted nucleon ($\approx 1-2$ MeV; see \citealt{haensel2008}) and $m_u$ is the atomic mass unit. 

We first note that, given the last recorded outburst in 2000 (Fig. \ref{fig:asm_lc}), and the dates of the {\it Chandra} (2013) and {\it XMM-Newton} (2018) observations, that shallow crustal heating is unlikely to explain the observed thermal luminosity. We then try to test the deep crustal heating scenario. Since we 
have some knowledge of 
the outburst history of our source (Fig. \ref{fig:asm_lc}), we can make a rough calculation of the average mass accretion rate and compare the inferred $L_q$ predicted by the deep crustal heating model with our observations. 
We extrapolated the {\it Chandra} $L_X$ down to 0.01 keV for bolometric correction, which gave us an $L_q$ of $1.23^{+1.66}_{-0.65}\times 10^{33}~\mathrm{erg~s^{-1}}$ (note that uncertainties in distance are not included in this result).

We estimate the average mass accretion rate of the source based on the observed $L_X$ during the outbursts,

\begin{equation}
    \langle L_X\rangle = 4\pi d^2 \langle F_X\rangle \sim \dfrac{GM\langle\dot M\rangle}{R}.
    \label{eq6}
\end{equation}
Using the folded light curve during the active epoch, we obtained a phase-averaged ASM count rate of $0.14\pm 0.06~\mathrm{counts/s}$. We convert this count rate to X-ray flux using the WebPIMMS tool\footnote{\url{https://heasarc.gsfc.nasa.gov/cgi-bin/Tools/w3pimms/w3pimms.pl}} and account for bolometric correction by assuming a power-law model between 0.1 keV and 12 keV, and adopting the best-fitting power-law index $\Gamma \sim 1$ and the e-folding energy $E_\mathrm{fold}\sim 12~\mathrm{keV}$ (for the upper limit of bolometric correction) from \cite{reig1999}. To account for uncertainty in the  bolometric correction due to the unknown spectral shape, we extended the power-law to 30 keV and calculated the combined error. The resulting ASM flux is $6.46^{+10.15}_{-2.54}\times 10^{-11}~\mathrm{erg~cm^{-2}~s^{-1}}$. If we only account for errors in the flux, and assume a 12-km NS with $M_\mathrm{NS}=1.4M_\odot$, we then have  $\langle \dot M \rangle_\mathrm{active} \approx 3.62^{+5.69}_{-1.42}\times 10^{-11}~M_\odot~\mathrm{yr^{-1}}$. This is only the accretion rate over the active period. To convert it to cover the whole period of ASM observations, we calculate the weighted mean,

\begin{equation}
\begin{aligned}
    \langle \dot M\rangle &= \dfrac{T_\mathrm{active} \langle \dot M\rangle_\mathrm{active} + T_\mathrm{quiescent} \langle \dot M\rangle_\mathrm{quiescent}}{T_\mathrm{active} + T_\mathrm{quiescent}} \\ 
    &\approx  \langle \dot M\rangle_\mathrm{active} \dfrac{T_\mathrm{active}}{T_\mathrm{active} + T_\mathrm{quiescent}} \\
    &\approx 5.82^{+9.16}_{-2.29}\times 10^{-12}~M_\odot\mathrm{~yr^{-1}},
\end{aligned}
    \label{eq7}
\end{equation}

where we have assumed that $\langle \dot M\rangle_\mathrm{active} \gg \langle \dot M\rangle_\mathrm{quiescent}$. 



\begin{figure}
    \centering
    \includegraphics[scale=0.42]{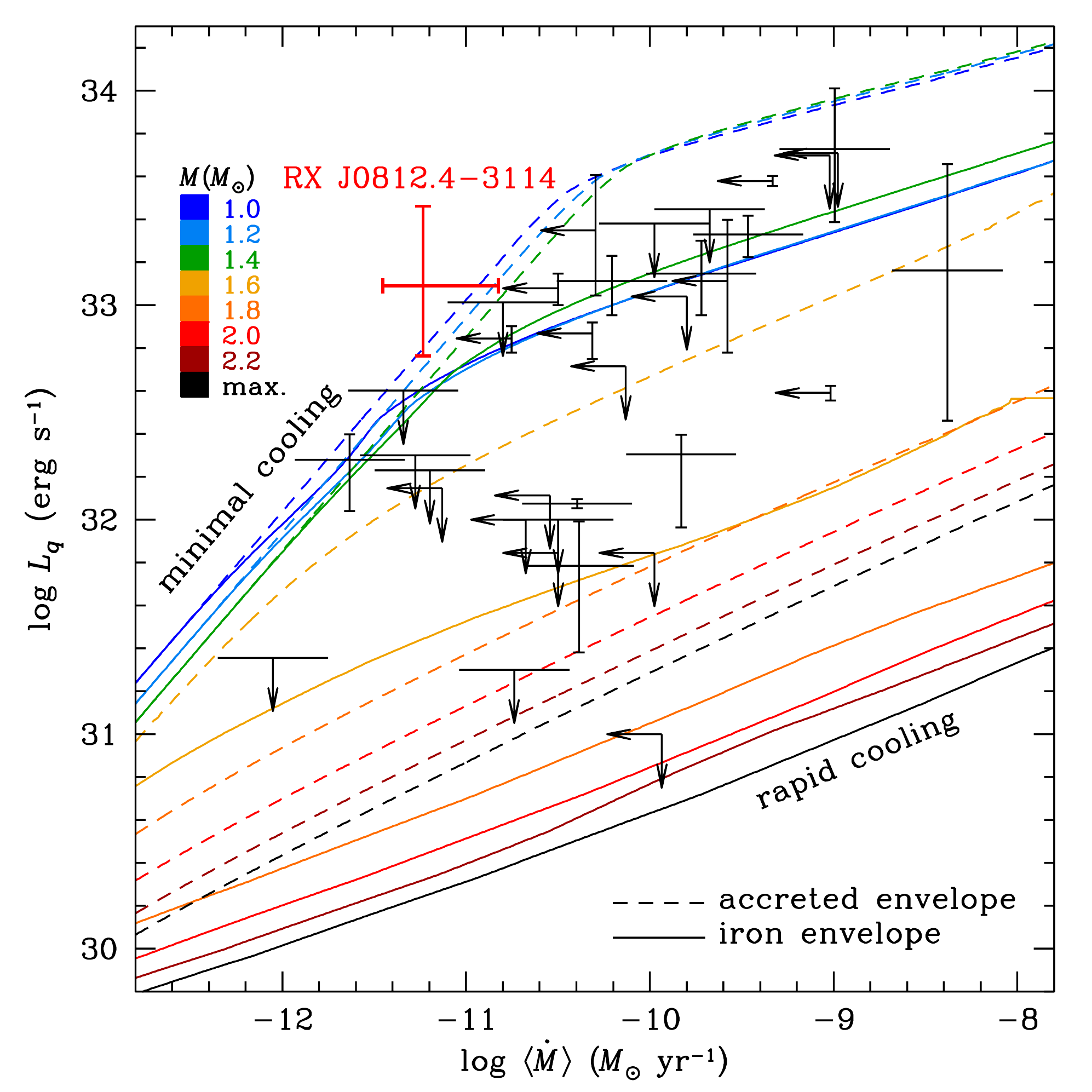}
    \caption{Observations of the quiescent thermal luminosities of low-mass X-ray
    binaries (qLMXB, black) and RX J0812.4$-$3114 (red), compared to
    theoretical predictions of the thermal luminosities (red-shifted as seen
    by a distant observer) produced by deep crustal heating for different
    time-averaged accretion rates. Theoretical models for neutron stars of
    different masses and different heat-blanketing envelope compositions
    (either iron or accreted helium and carbon), as well as the qLMXB data,
    are taken from Potekhin, Chugunov, \& Chabrier (2019, submitted). The
    relatively low-mass neutron stars in qLMXBs (upper curves) undergo the
    minimal cooling (e.g., \citealt{Page04}), whereas the high-mass neutron
    stars (lower curves) undergo rapid cooling via direct Urca process. The
    observational  accretion rates are scaled from the canonical $R=10$~km
    in Potekhin et al. to the more probable $R=12$~km in the present paper.
    RX J0812.4$-$3114 lies at or above the minimal cooling tracks.}
    \label{fig:lq_vs_mdot}
\end{figure}

\red{To compare with other quiescent systems, we plot observed quiescent NSs in LMXBs on a $L_q-\langle \dot M\rangle$ plot (Fig. \ref{fig:lq_vs_mdot}) with tracks for possible cooling mechanisms indicated. The theoretical tracks are calculated assuming the BSk24 equation of state \citep{pearson2018} with a maximum mass of 2.28$M_\odot$ and a mass threshold for rapid cooling of 1.595$M_\odot$; for modified-Urca processes, we included in-medium effect following \cite{shternin2018}. We used the MSH and BS gap models from \cite{Ho2015} and accounted for effects of triplet superfluidity following \cite{ding2016}. J0812 likely lies above the minimum cooling curves for NSs with iron heat-blanketing envelopes (for a definition of heat-blanketing envelope, see \citealt{gudmundsson1983}), but is consistent with low-mass (1.0-1.2 $M_\odot$) NSs with accreted heat-blanketing envelopes. 

J0812's position above some of the minimum cooling tracks may require explanation. 
We speculate that this NS might be relatively young, so that the NS has not yet lost the internal heat deposited in its supernova explosion. Reference to, e.g., \citet{Page04} shows that the thermal $L_X$ from the supernova should fall below $\sim 10^{33}$ erg/s after $10^5$ years. J0812's companion's B0 spectral type indicates a $\sim 20 M_\odot$ mass, and thus a $\lesssim 2$ Myr lifetime, suggesting a $\gtrsim 5$\% chance of observing a NS in this HMXB before it has lost its supernova heat. }


We can attempt to estimate the age since the SN from its height above the Galactic Plane, and proper motion. 
J0812 would take $10^8$ years, at its Gaia-measured proper motion \citep{Gaia18}, to reach its location $\approx 1\fdg54$ above the Galactic Plane. However, considering the much shorter lifetime of the companion star, it was almost certainly born near its current location, and this method cannot constrain this hypothesis.

\subsection{Hard component: stable low-level accretion}
The hard (power-law) component in quiescent accreting NSs is generally thought to originate from low-level accretion \citep[e.g.][]{wijnands2015}. In high-B systems, the accretion columns are channeled down to the magnetic poles, so the hard component is likely to be pulsed (e.g., in Cep X-4; see \citealt{mcbride2007}). Theoretically, accretion onto the NS poles is only possible when the rotational velocity of the magnetic field lines is lower than the local Keplerian velocity. Otherwise, matter from the accretion disc would be spun away by the centrifugal barrier, making the system enter the propeller regime. Observationally, the onset of the propeller regime is marked by an abrupt drop in accretion luminosity ($L_\mathrm{acc}$) when the luminosity decays below a limiting level ($L_\mathrm{prop}$). The limiting luminosity is estimated by equating the magnetospheric radius ($R_\mathrm{m}$) with the corotation radius ($R_\mathrm{c}$), where the local Keplerian velocity equals the rotational velocity of the NS. One can then derive the following\footnote{Notice that the magnetic field here is the dipolar strength at the magnetic poles, which is a factor of 2 higher than that at the equator.}

\begin{equation}
    L_\mathrm{prop} \approx 4\times 10^{37} \xi^{7/2} B_{12}^2 P^{-7/3} M_{1.4}^{-2/3} R_6^5~\mathrm{erg~s^{-1}},
    \label{eq8}
\end{equation}
where $\xi$ is a parameter that defines the accretion geometry ($\xi=0.5$ for accretion from a disc; $\xi=1$ for accretion from winds; \citealt{Ghosh78}); $B_{12}$ is the magnetic field strength in units of $10^{12}~\mathrm{G}$; $P$ is the spin period of the NS; $M_{1.4}$ is the mass of the NS in units of $1.4M_\odot$, and $R_6$ is the NS radius in units of $10^6~\mathrm{cm}$. 

We do not know $L_\mathrm{prop}$ for J0812, its magnetic field is unknown. However, it is plausible to assume that the source was in the propeller regime ($L_\mathrm{acc} \lesssim L_\mathrm{prop}$) during the 2013 {\it Chandra} observation, based on the fact that the source possessed a very soft spectrum (consistent with emission of stored heat from the NS with no active accretion), and assume that the source in 2018 had left the propeller regime and was accreting 
($L_\mathrm{acc} \gtrsim L_\mathrm{prop}$), based on the fact that a pulsed hard power-law component is present. We estimate the 2018 $L_\mathrm{acc}$ by using the power-law component ($\Gamma=1.49$) from the {\tt gabs*(pow+nsmaxg)} \# 2 fit and extrapolate it to 0.1-30 keV for bolometric correction. This yields an $L_\mathrm{0.1-30}=1.80\times 10^{33}~\mathrm{erg~s^{-1}}$, which is then adopted as an upper limit on the propeller luminosity ($L_\mathrm{prop}$). With this constraint on $L_\mathrm{prop}$, we can then place an upper limit on the magnetic field strength of the NS using eq.(\ref{eq8}). 
We assumed an NS of $1.4M_\odot$ and $R_\mathrm{NS}=12~\mathrm{km}$. 
J0812 has been quiescent for many years, so it might seem plausible to assume accretion from the winds of the Be star (i.e. $\xi=1$). However, a disc may still form in even wind-fed systems \citep{Karino19} or systems fed by a companion star's circumstellar disc \citep{Klus14}. We therefore calculated for cases of $\xi=0.5$ and $\xi=1$ and found $B_{12}(\xi=1)$ $\lesssim 0.24$ while $B_{12}(\xi=0.5)=0.81$, corresponding to cyclotron energies $\lesssim 2.81~\mathrm{keV}$ and $\lesssim 9.44~\mathrm{keV}$, respectively. A recent study by \citet{campana2018} on different classes of objects subject to the propeller effect found a best fit with $\xi = 0.49 \pm 0.05$, which results in a different scaling factor of $\approx 3.3\times10^{36}~\mathrm{erg~s^{-1}}$ for eq.(\ref{eq8})\footnote{This scaling factor is different from eq.(7) in \citet{campana2018} because we write $M$ in units of $1.4M_\odot$ and use $B$ as the field strength at the  magnetic poles, to be consistent with eq. (\ref{eq8})}. Using this factor, we obtained $B_\mathrm{12}\lesssim 0.84$ and cyclotron energies $\lesssim 9.78~\mathrm{keV}$. It should be noted that this is a rough estimate on the magnetic field, subject to errors in distance and fluxes, 
but the possible absorption line, \red{either the one at $1~\mathrm{keV}$ indicated from the {\tt gabs*(pow+bbodyrad)} fit or the one at $1.4$ keV suggested by the {\tt gabs*(pow+nsmaxg)} fit}, might then be real, as the inferred upper limit from the propeller argument (in the case of $\xi=1$) is very close to what we found in the spectrum. Therefore the inferred $B$-field from this line, $\sim 10^{11}$ G, might 
be the magnetic field of this system.
Verification of this relatively low $B$ field for a HMXB NS could be confirmed by future \blue{high-sensitivity observations that cover a broad energy range, e.g. observations with NICER and NuSTAR during outburst.}

\section{Conclusion}
Our 72 ks {\it XMM-Newton} observation of the quiescent BeXRP RX J0812.4-3114 revealed a 
spectrum best described by a soft blackbody component plus a hard power-law, with an absorption component at $\approx 1~\mathrm{keV}$ \red{and/or at $\approx 1.4~\mathrm{keV}$}. The blackbody component implies a large emission region, which we argue likely originates from the whole NS surface. The hard component can be described by a hard power-law  ($\Gamma\sim 1.3-1.5$), likely caused by low-level accretion. 

Temporal analyses reveal that the hard component is pulsed at a period of $31.908\pm 0.009~\mathrm{s}$, slightly longer than the previous studies, from which we estimated a spin-down rate of $\approx 3.63\times 10^{-11}~\mathrm{s~s^{-1}}$, which might just be due to orbital Doppler effect. We did not find pulsations in the soft excess ($\mathrm{PF}\lesssim 31\%$), 
indicating 
the soft emission has a different origin from the hard emission (consistent with emission from the full NS surface). 
Long-term lightcurves reveal variability in the hard component; however no sign of variability was found in the soft excess. Temporal analysis of the ASM light curves confirms the previously measured orbital period of $\approx 81.3$ days, and that the source returned to a quiescent state after the active epoch between 1997-12-05 and 2000-07-02. Based on the accretion history, we estimated the time-averaged mass accretion rate ($\langle \dot M\rangle$). Assuming the quiescent thermal luminosity is produced by deep crustal heating, we 
found that J0812 lies above some of the minimum cooling tracks, though with large uncertainty in $\langle \dot M\rangle$. The NS in J0812 may yet agree with minimum cooling processes. However, it is also possible that the NS in J0812 is too young to have fully cooled after its supernova explosion -- a possibility which we estimate to have a $\gtrsim 5$\% chance, given the estimated lifetime of the B0 companion and the timescale for NS cooling. 


J0812 seems to have two distinct X-ray spectral states in quiescence: a soft, or thermally dominated state as observed by {\it Chandra}, versus a harder state with possible on-going accretion as observed by our {\it XMM-Newton} observation. We suspect the source lies in the propeller regime during the soft state, while 
it is out of the propeller regime and accreting during the hard state. 
With these assumptions, we estimate the magnetic field strength of the system to be $\lesssim 8.4 \times10^{11}~\mathrm{G}$. 
Should the $\approx$1 keV \red{or $\approx 1.4$ keV} absorption feature be real, and represent an electron cyclotron line, then we may further estimate $B\sim10^{11}$ G, unusually low for BeXRPs, but consistent with the estimate from the propeller arguement.

\label{sec:conclusion}














\section*{Acknowledgements}
COH is supported by NSERC Discovery Grant RGPIN-2016-04602 and a Discovery Accelerator Supplement. \blue{COH thanks the organizers and participants of the April 2019 "Investigating Crusts Of Neutron Stars" workshop at the Anton Pannekoek Institute of the University of Amsterdam, supported by JINA-CEE, for discussions.} SST was supported by the grant 14.W03.31.0021 of the Ministry of Science and Higher Education of the Russian Federation. WCGH appreciates use of computer facilities at the Kavli Institute for Particle Astrophysics and Cosmology. The work of AYP was supported by RFBR and DFG within the research
project 19-52-12013. This work is primarily based on observations obtained with {it XMM-Newton}, an ESA science mission with instruments and contributions directly funded by ESA Member States and NASA. We also use data from the European Space Agency (ESA) mission {\it Gaia} (\url{https://www.cosmos.esa.int/gaia}), processed by the {\it Gaia}
Data Processing and Analysis Consortium (DPAC,
\url{https://www.cosmos.esa.int/web/gaia/dpac/consortium}). Funding for the DPAC
has been provided by national institutions, in particular the institutions
participating in the {\it Gaia} Multilateral Agreement. Spectral and temporal analyses in this work made use of data and/or software provided by the High Energy Astrophysics Science Archive Research Center (HEASARC), which is a service of the Astrophysics Science Division at NASA/GSFC and the High Energy Astrophysics Division of the Smithsonian Astrophysical Observatory. This research has also made use of the NASA Astrophysics Data System (ADS) and software provided by the Chandra X-ray Center (CXC) in the application package {\sc CIAO}. 




\bibliographystyle{mnras}

\begin{thebibliography}{}
\makeatletter
\relax
\def\mn@urlcharsother{\let\do\@makeother \do\$\do\&\do\#\do\^\do\_\do\%\do\~}
\def\mn@doi{\begingroup\mn@urlcharsother \@ifnextchar [ {\mn@doi@}
  {\mn@doi@[]}}
\def\mn@doi@[#1]#2{\def\@tempa{#1}\ifx\@tempa\@empty \href
  {http://dx.doi.org/#2} {doi:#2}\else \href {http://dx.doi.org/#2} {#1}\fi
  \endgroup}
\def\mn@eprint#1#2{\mn@eprint@#1:#2::\@nil}
\def\mn@eprint@arXiv#1{\href {http://arxiv.org/abs/#1} {{\tt arXiv:#1}}}
\def\mn@eprint@dblp#1{\href {http://dblp.uni-trier.de/rec/bibtex/#1.xml}
  {dblp:#1}}
\def\mn@eprint@#1:#2:#3:#4\@nil{\def\@tempa {#1}\def\@tempb {#2}\def\@tempc
  {#3}\ifx \@tempc \@empty \let \@tempc \@tempb \let \@tempb \@tempa \fi \ifx
  \@tempb \@empty \def\@tempb {arXiv}\fi \@ifundefined
  {mn@eprint@\@tempb}{\@tempb:\@tempc}{\expandafter \expandafter \csname
  mn@eprint@\@tempb\endcsname \expandafter{\@tempc}}}

\bibitem[\protect\citeauthoryear{{Bailer-Jones}, {Rybizki}, {Fouesneau},
  {Mantelet}  \& {Andrae}}{{Bailer-Jones} et~al.}{2018}]{bailer2018}
{Bailer-Jones} C.~A.~L.,  {Rybizki} J.,  {Fouesneau} M.,  {Mantelet} G.,
  {Andrae} R.,  2018, \mn@doi [\aj] {10.3847/1538-3881/aacb21}, \href
  {https://ui.adsabs.harvard.edu/#abs/2018AJ....156...58B} {156, 58}

\bibitem[\protect\citeauthoryear{{Brown} \& {Cumming}}{{Brown} \&
  {Cumming}}{2009}]{Brown09}
{Brown} E.~F.,  {Cumming} A.,  2009, \mn@doi [\apj]
  {10.1088/0004-637X/698/2/1020}, \href
  {http://adsabs.harvard.edu/abs/2009ApJ...698.1020B} {698, 1020}

\bibitem[\protect\citeauthoryear{{Brown}, {Bildsten}  \& {Rutledge}}{{Brown}
  et~al.}{1998}]{brown1998}
{Brown} E.~F.,  {Bildsten} L.,   {Rutledge} R.~E.,  1998, \mn@doi [\apj]
  {10.1086/311578}, \href
  {https://ui.adsabs.harvard.edu/\#abs/1998ApJ...504L..95B} {504, L95}

\bibitem[\protect\citeauthoryear{{Burderi}, {Di Salvo}, {Robba}, {La Barbera}
  \& {Guainazzi}}{{Burderi} et~al.}{2000}]{burderi2000}
{Burderi} L.,  {Di Salvo} T.,  {Robba} N.~R.,  {La Barbera} A.,   {Guainazzi}
  M.,  2000, \mn@doi [\apj] {10.1086/308336}, \href
  {https://ui.adsabs.harvard.edu/\#abs/2000ApJ...530..429B} {530, 429}

\bibitem[\protect\citeauthoryear{{Campana}, {Gastaldello}, {Stella}, {Israel},
  {Colpi}, {Pizzolato}, {Orlandini}  \& {Dal Fiume}}{{Campana}
  et~al.}{2001}]{campana01}
{Campana} S.,  {Gastaldello} F.,  {Stella} L.,  {Israel} G.~L.,  {Colpi} M.,
  {Pizzolato} F.,  {Orlandini} M.,   {Dal Fiume} D.,  2001, \mn@doi [\apj]
  {10.1086/323317}, \href {http://adsabs.harvard.edu/abs/2001ApJ...561..924C}
  {561, 924}

\bibitem[\protect\citeauthoryear{{Campana}, {Stella}, {Israel}, {Moretti},
  {Parmar}  \& {Orlandini}}{{Campana} et~al.}{2002}]{campana2002}
{Campana} S.,  {Stella} L.,  {Israel} G.~L.,  {Moretti} A.,  {Parmar} A.~N.,
  {Orlandini} M.,  2002, \mn@doi [\apj] {10.1086/343074}, \href
  {https://ui.adsabs.harvard.edu/\#abs/2002ApJ...580..389C} {580, 389}

\bibitem[\protect\citeauthoryear{{Campana}, {Stella}, {Mereghetti}  \& {de
  Martino}}{{Campana} et~al.}{2018}]{campana2018}
{Campana} S.,  {Stella} L.,  {Mereghetti} S.,   {de Martino} D.,  2018, \mn@doi
  [\aap] {10.1051/0004-6361/201730769}, \href
  {https://ui.adsabs.harvard.edu/abs/2018A&A...610A..46C} {610, A46}

\bibitem[\protect\citeauthoryear{{Cash}}{{Cash}}{1979}]{cash1979}
{Cash} W.,  1979, \mn@doi [\apj] {10.1086/156922}, \href
  {http://adsabs.harvard.edu/abs/1979ApJ...228..939C} {228, 939}

\bibitem[\protect\citeauthoryear{{Coleiro} \& {Chaty}}{{Coleiro} \&
  {Chaty}}{2013}]{coleiro2013}
{Coleiro} A.,  {Chaty} S.,  2013, \mn@doi [\apj] {10.1088/0004-637X/764/2/185},
  \href {https://ui.adsabs.harvard.edu/abs/2013ApJ...764..185C} {764, 185}

\bibitem[\protect\citeauthoryear{{Corbet}}{{Corbet}}{1999}]{corbet1999}
{Corbet} R.,  1999, International Astronomical Union Circular, \href
  {https://ui.adsabs.harvard.edu/\#abs/1999IAUC.7104....2C} {7104, 2}

\bibitem[\protect\citeauthoryear{{Corbet} \& {Peele}}{{Corbet} \&
  {Peele}}{2000}]{corbet2000}
{Corbet} R.~H.~D.,  {Peele} A.~G.,  2000, \mn@doi [\apjl] {10.1086/312485},
  \href {http://adsabs.harvard.edu/abs/2000ApJ...530L..33C} {530, L33}

\bibitem[\protect\citeauthoryear{{Deibel}, {Cumming}, {Brown}  \&
  {Page}}{{Deibel} et~al.}{2015}]{deibel2015}
{Deibel} A.,  {Cumming} A.,  {Brown} E.~F.,   {Page} D.,  2015, \mn@doi [\apj]
  {10.1088/2041-8205/809/2/L31}, \href
  {https://ui.adsabs.harvard.edu/\#abs/2015ApJ...809L..31D} {809, L31}

\bibitem[\protect\citeauthoryear{{Ding}, {Rios}, {Dussan}, {Dickhoff}, {Witte},
  {Carbone}  \& {Polls}}{{Ding} et~al.}{2016}]{ding2016}
{Ding} D.,  {Rios} A.,  {Dussan} H.,  {Dickhoff} W.~H.,  {Witte} S.~J.,
  {Carbone} A.,   {Polls} A.,  2016, \mn@doi [\prc]
  {10.1103/PhysRevC.94.025802}, \href
  {https://ui.adsabs.harvard.edu/abs/2016PhRvC..94b5802D} {94, 025802}

\bibitem[\protect\citeauthoryear{{Doroshenko}, {Santangelo}  \&
  {Suleimanov}}{{Doroshenko} et~al.}{2011}]{Doroshenko11}
{Doroshenko} V.,  {Santangelo} A.,   {Suleimanov} V.,  2011, \mn@doi [\aap]
  {10.1051/0004-6361/201116482}, \href
  {http://adsabs.harvard.edu/abs/2011A%26A...529A..52D} {529, A52}

\bibitem[\protect\citeauthoryear{{Doroshenko}, {Santangelo}, {Doroshenko},
  {Caballero}, {Tsygankov}  \& {Rothschild}}{{Doroshenko}
  et~al.}{2014}]{Doroshenko14}
{Doroshenko} V.,  {Santangelo} A.,  {Doroshenko} R.,  {Caballero} I.,
  {Tsygankov} S.,   {Rothschild} R.,  2014, \mn@doi [\aap]
  {10.1051/0004-6361/201322472}, \href
  {http://adsabs.harvard.edu/abs/2014A%26A...561A..96D} {561, A96}

\bibitem[\protect\citeauthoryear{{Elshamouty}, {Heinke}, {Morsink}, {Bogdanov}
  \& {Stevens}}{{Elshamouty} et~al.}{2016}]{elshamouty2016}
{Elshamouty} K.~G.,  {Heinke} C.~O.,  {Morsink} S.~M.,  {Bogdanov} S.,
  {Stevens} A.~L.,  2016, \mn@doi [\apj] {10.3847/0004-637X/826/2/162}, \href
  {https://ui.adsabs.harvard.edu/#abs/2016ApJ...826..162E} {826, 162}

\bibitem[\protect\citeauthoryear{{Endo}, {Nagase}  \& {Mihara}}{{Endo}
  et~al.}{2000}]{endo2000}
{Endo} T.,  {Nagase} F.,   {Mihara} T.,  2000, \mn@doi [Publications of the
  Astronomical Society of Japan] {10.1093/pasj/52.2.223}, \href
  {https://ui.adsabs.harvard.edu/\#abs/2000PASJ...52..223E} {52, 223}

\bibitem[\protect\citeauthoryear{{Forestell}, {Heinke}, {Cohn}, {Lugger},
  {Sivakoff}, {Bogdanov}, {Cool}  \& {Anderson}}{{Forestell}
  et~al.}{2014}]{forestell2014}
{Forestell} L.~M.,  {Heinke} C.~O.,  {Cohn} H.~N.,  {Lugger} P.~M.,  {Sivakoff}
  G.~R.,  {Bogdanov} S.,  {Cool} A.~M.,   {Anderson} J.,  2014, \mn@doi
  [\mnras] {10.1093/mnras/stu559}, \href
  {http://adsabs.harvard.edu/abs/2014MNRAS.441..757F} {441, 757}

\bibitem[\protect\citeauthoryear{{Gaia Collaboration} et~al.,}{{Gaia
  Collaboration} et~al.}{2018}]{Gaia18}
{Gaia Collaboration} et~al., 2018, \mn@doi [\aap]
  {10.1051/0004-6361/201833051}, \href
  {http://adsabs.harvard.edu/abs/2018A%26A...616A...1G} {616, A1}

\bibitem[\protect\citeauthoryear{{Galloway}, {Giles}, {Wu}  \&
  {Greenhill}}{{Galloway} et~al.}{2001}]{galloway2001}
{Galloway} D.~K.,  {Giles} A.~B.,  {Wu} K.,   {Greenhill} J.~G.,  2001, \mn@doi
  [\mnras] {10.1046/j.1365-8711.2001.04467.x}, \href
  {http://adsabs.harvard.edu/abs/2001MNRAS.325..419G} {325, 419}

\bibitem[\protect\citeauthoryear{{Geppert}, {K{\"u}ker}  \& {Page}}{{Geppert}
  et~al.}{2004}]{geppert2004}
{Geppert} U.,  {K{\"u}ker} M.,   {Page} D.,  2004, \mn@doi [\aap]
  {10.1051/0004-6361:20040455}, \href
  {https://ui.adsabs.harvard.edu/#abs/2004A&A...426..267G} {426, 267}

\bibitem[\protect\citeauthoryear{{Ghosh} \& {Lamb}}{{Ghosh} \&
  {Lamb}}{1978}]{Ghosh78}
{Ghosh} P.,  {Lamb} F.~K.,  1978, \mn@doi [\apjl] {10.1086/182734}, \href
  {https://ui.adsabs.harvard.edu/abs/1978ApJ...223L..83G} {223, L83}

\bibitem[\protect\citeauthoryear{{Gotthelf} \& {Halpern}}{{Gotthelf} \&
  {Halpern}}{2009}]{gotthelf2009}
{Gotthelf} E.~V.,  {Halpern} J.~P.,  2009, \mn@doi [\apj]
  {10.1088/0004-637X/695/1/L35}, \href
  {https://ui.adsabs.harvard.edu/#abs/2009ApJ...695L..35G} {695, L35}

\bibitem[\protect\citeauthoryear{{Greenstein}, {Dolez}  \&
  {Vauclair}}{{Greenstein} et~al.}{1983}]{greenstein1983}
{Greenstein} J.~L.,  {Dolez} N.,   {Vauclair} G.,  1983, \aap, \href
  {https://ui.adsabs.harvard.edu/#abs/1983A&A...127...25G} {127, 25}

\bibitem[\protect\citeauthoryear{{Gudmundsson}, {Pethick}  \&
  {Epstein}}{{Gudmundsson} et~al.}{1983}]{gudmundsson1983}
{Gudmundsson} E.~H.,  {Pethick} C.~J.,   {Epstein} R.~I.,  1983, \mn@doi [\apj]
  {10.1086/161292}, \href
  {https://ui.adsabs.harvard.edu/abs/1983ApJ...272..286G} {272, 286}

\bibitem[\protect\citeauthoryear{{Haensel} \& {Zdunik}}{{Haensel} \&
  {Zdunik}}{2008}]{haensel2008}
{Haensel} P.,  {Zdunik} J.~L.,  2008, \mn@doi [\aap]
  {10.1051/0004-6361:20078578}, \href
  {https://ui.adsabs.harvard.edu/\#abs/2008A&A...480..459H} {480, 459}

\bibitem[\protect\citeauthoryear{{Hickox}, {Narayan}  \& {Kallman}}{{Hickox}
  et~al.}{2004}]{hickox2004}
{Hickox} R.~C.,  {Narayan} R.,   {Kallman} T.~R.,  2004, \mn@doi [\apj]
  {10.1086/423928}, \href {http://adsabs.harvard.edu/abs/2004ApJ...614..881H}
  {614, 881}

\bibitem[\protect\citeauthoryear{{Ho}, {Potekhin}  \& {Chabrier}}{{Ho}
  et~al.}{2008}]{ho2008}
{Ho} W.~C.~G.,  {Potekhin} A.~Y.,   {Chabrier} G.,  2008, \mn@doi [\apjs]
  {10.1086/589238}, \href {http://adsabs.harvard.edu/abs/2008ApJS..178..102H}
  {178, 102}

\bibitem[\protect\citeauthoryear{{Ho}, {Elshamouty}, {Heinke}  \&
  {Potekhin}}{{Ho} et~al.}{2015}]{Ho2015}
{Ho} W. C.~G.,  {Elshamouty} K.~G.,  {Heinke} C.~O.,   {Potekhin} A.~Y.,  2015,
  \mn@doi [\prc] {10.1103/PhysRevC.91.015806}, \href
  {https://ui.adsabs.harvard.edu/abs/2015PhRvC..91a5806H} {91, 015806}

\bibitem[\protect\citeauthoryear{{Illarionov} \& {Sunyaev}}{{Illarionov} \&
  {Sunyaev}}{1975}]{illarionov1975}
{Illarionov} A.~F.,  {Sunyaev} R.~A.,  1975, \aap, \href
  {http://adsabs.harvard.edu/abs/1975A%26A....39..185I} {39, 185}

\bibitem[\protect\citeauthoryear{{Kalberla}, {Burton}, {Hartmann}, {Arnal},
  {Bajaja}, {Morras}  \& {P{\"o}ppel}}{{Kalberla} et~al.}{2005}]{kalberla2005}
{Kalberla} P.~M.~W.,  {Burton} W.~B.,  {Hartmann} D.,  {Arnal} E.~M.,  {Bajaja}
  E.,  {Morras} R.,   {P{\"o}ppel} W.~G.~L.,  2005, \mn@doi [\aap]
  {10.1051/0004-6361:20041864}, \href
  {http://adsabs.harvard.edu/abs/2005A%26A...440..775K} {440, 775}

\bibitem[\protect\citeauthoryear{{Karino}, {Nakamura}  \& {Taani}}{{Karino}
  et~al.}{2019}]{Karino19}
{Karino} S.,  {Nakamura} K.,   {Taani} A.,  2019, \mn@doi [\pasj]
  {10.1093/pasj/psz034}, \href
  {https://ui.adsabs.harvard.edu/abs/2019PASJ...71...58K} {71, 58}

\bibitem[\protect\citeauthoryear{{Klus}, {Ho}, {Coe}, {Corbet}  \&
  {Townsend}}{{Klus} et~al.}{2014}]{Klus14}
{Klus} H.,  {Ho} W.~C.~G.,  {Coe} M.~J.,  {Corbet} R.~H.~D.,   {Townsend}
  L.~J.,  2014, \mn@doi [\mnras] {10.1093/mnras/stt2192}, \href
  {https://ui.adsabs.harvard.edu/abs/2014MNRAS.437.3863K} {437, 3863}

\bibitem[\protect\citeauthoryear{{La Palombara} \& {Mereghetti}}{{La Palombara}
  \& {Mereghetti}}{2006}]{LaPalombara06}
{La Palombara} N.,  {Mereghetti} S.,  2006, \mn@doi [\aap]
  {10.1051/0004-6361:20065107}, \href
  {http://adsabs.harvard.edu/abs/2006A%26A...455..283L} {455, 283}

\bibitem[\protect\citeauthoryear{{La Palombara} \& {Mereghetti}}{{La Palombara}
  \& {Mereghetti}}{2007}]{LaPalombara07}
{La Palombara} N.,  {Mereghetti} S.,  2007, \mn@doi [\aap]
  {10.1051/0004-6361:20077970}, \href
  {http://adsabs.harvard.edu/abs/2007A%26A...474..137L} {474, 137}

\bibitem[\protect\citeauthoryear{{La Palombara} \& {Mereghetti}}{{La Palombara}
  \& {Mereghetti}}{2017}]{palombara2017}
{La Palombara} N.,  {Mereghetti} S.,  2017, \mn@doi [\aap]
  {10.1051/0004-6361/201730887}, \href
  {https://ui.adsabs.harvard.edu/\#abs/2017A&A...602A.114L} {602, A114}

\bibitem[\protect\citeauthoryear{{La Palombara}, {Sidoli}, {Esposito}, {Tiengo}
   \& {Mereghetti}}{{La Palombara} et~al.}{2009a}]{LaPalombara09}
{La Palombara} N.,  {Sidoli} L.,  {Esposito} P.,  {Tiengo} A.,   {Mereghetti}
  S.,  2009a, \mn@doi [\aap] {10.1051/0004-6361/200912538}, \href
  {http://adsabs.harvard.edu/abs/2009A%26A...505..947L} {505, 947}

\bibitem[\protect\citeauthoryear{{La Palombara}, {Sidoli}, {Esposito}, {Tiengo}
   \& {Mereghetti}}{{La Palombara} et~al.}{2009b}]{palombara2009}
{La Palombara} N.,  {Sidoli} L.,  {Esposito} P.,  {Tiengo} A.,   {Mereghetti}
  S.,  2009b, \mn@doi [\aap] {10.1051/0004-6361/200912538}, \href
  {https://ui.adsabs.harvard.edu/\#abs/2009A&A...505..947L} {505, 947}

\bibitem[\protect\citeauthoryear{{Leahy}, {Darbro}, {Elsner}, {Weisskopf},
  {Sutherland}, {Kahn}  \& {Grindlay}}{{Leahy} et~al.}{1983}]{Leahy83}
{Leahy} D.~A.,  {Darbro} W.,  {Elsner} R.~F.,  {Weisskopf} M.~C.,  {Sutherland}
  P.~G.,  {Kahn} S.,   {Grindlay} J.~E.,  1983, \mn@doi [\apj]
  {10.1086/160766}, \href
  {https://ui.adsabs.harvard.edu/abs/1983ApJ...266..160L} {266, 160}

\bibitem[\protect\citeauthoryear{{Lyne} \& {Graham-Smith}}{{Lyne} \&
  {Graham-Smith}}{2006}]{LyneGrahamSmith2006}
{Lyne} A.~G.,  {Graham-Smith} F.,  2006, {Pulsar Astronomy}

\bibitem[\protect\citeauthoryear{{McBride} et~al.,}{{McBride}
  et~al.}{2007}]{mcbride2007}
{McBride} V.~A.,  et~al., 2007, \mn@doi [\aap] {10.1051/0004-6361:20077238},
  \href {https://ui.adsabs.harvard.edu/#abs/2007A&A...470.1065M} {470, 1065}

\bibitem[\protect\citeauthoryear{{Motch} et~al.,}{{Motch}
  et~al.}{1991}]{motch1991}
{Motch} C.,  et~al., 1991, \aap, \href
  {http://adsabs.harvard.edu/abs/1991A%26A...246L..24M} {246, L24}

\bibitem[\protect\citeauthoryear{{Motch}, {Haberl}, {Dennerl}, {Pakull}  \&
  {Janot-Pacheco}}{{Motch} et~al.}{1997}]{motch1997}
{Motch} C.,  {Haberl} F.,  {Dennerl} K.,  {Pakull} M.,   {Janot-Pacheco} E.,
  1997, \aap, \href {http://adsabs.harvard.edu/abs/1997A%26A...323..853M} {323,
  853}

\bibitem[\protect\citeauthoryear{{Mukherjee} \& {Paul}}{{Mukherjee} \&
  {Paul}}{2005}]{mukherjee2005}
{Mukherjee} U.,  {Paul} B.,  2005, \mn@doi [\aap] {10.1051/0004-6361:20041665},
  \href {http://adsabs.harvard.edu/abs/2005A%26A...431..667M} {431, 667}

\bibitem[\protect\citeauthoryear{{Naz{\'e}} et~al.,}{{Naz{\'e}}
  et~al.}{2011}]{naze2011}
{Naz{\'e}} Y.,  et~al., 2011, \mn@doi [The Astrophysical Journal Supplement
  Series] {10.1088/0067-0049/194/1/7}, \href
  {https://ui.adsabs.harvard.edu/#abs/2011ApJS..194....7N} {194, 7}

\bibitem[\protect\citeauthoryear{{Negueruela}, {Reig}, {Finger}  \&
  {Roche}}{{Negueruela} et~al.}{2000}]{negueruela2000}
{Negueruela} I.,  {Reig} P.,  {Finger} M.~H.,   {Roche} P.,  2000, \aap, \href
  {https://ui.adsabs.harvard.edu/\#abs/2000A&A...356.1003N} {356, 1003}

\bibitem[\protect\citeauthoryear{{Page}, {Lattimer}, {Prakash}  \&
  {Steiner}}{{Page} et~al.}{2004}]{Page04}
{Page} D.,  {Lattimer} J.~M.,  {Prakash} M.,   {Steiner} A.~W.,  2004, \mn@doi
  [\apjs] {10.1086/424844}, \href
  {http://adsabs.harvard.edu/abs/2004ApJS..155..623P} {155, 623}

\bibitem[\protect\citeauthoryear{{Pearson}, {Chamel}, {Potekhin}, {Fantina},
  {Ducoin}, {Dutta}  \& {Goriely}}{{Pearson} et~al.}{2018}]{pearson2018}
{Pearson} J.~M.,  {Chamel} N.,  {Potekhin} A.~Y.,  {Fantina} A.~F.,  {Ducoin}
  C.,  {Dutta} A.~K.,   {Goriely} S.,  2018, \mn@doi [\mnras]
  {10.1093/mnras/sty2413}, \href
  {https://ui.adsabs.harvard.edu/abs/2018MNRAS.481.2994P} {481, 2994}

\bibitem[\protect\citeauthoryear{{Potekhin} \& {Yakovlev}}{{Potekhin} \&
  {Yakovlev}}{2001}]{potekhin2001}
{Potekhin} A.~Y.,  {Yakovlev} D.~G.,  2001, \mn@doi [\aap]
  {10.1051/0004-6361:20010698}, \href
  {https://ui.adsabs.harvard.edu/#abs/2001A&A...374..213P} {374, 213}

\bibitem[\protect\citeauthoryear{{Potekhin}, {Chabrier}  \& {Ho}}{{Potekhin}
  et~al.}{2014}]{potekhin2014}
{Potekhin} A.~Y.,  {Chabrier} G.,   {Ho} W.~C.~G.,  2014, \mn@doi [\aap]
  {10.1051/0004-6361/201424619}, \href
  {https://ui.adsabs.harvard.edu/abs/2014A&A...572A..69P} {572, A69}

\bibitem[\protect\citeauthoryear{{Potekhin}, {Pons}  \& {Page}}{{Potekhin}
  et~al.}{2015}]{potekhin2015}
{Potekhin} A.~Y.,  {Pons} J.~A.,   {Page} D.,  2015, \mn@doi [\ssr]
  {10.1007/s11214-015-0180-9}, \href
  {https://ui.adsabs.harvard.edu/abs/2015SSRv..191..239P} {191, 239}

\bibitem[\protect\citeauthoryear{{Reig}}{{Reig}}{2011}]{reig2011}
{Reig} P.,  2011, \mn@doi [\apss] {10.1007/s10509-010-0575-8}, \href
  {http://adsabs.harvard.edu/abs/2011Ap%26SS.332....1R} {332, 1}

\bibitem[\protect\citeauthoryear{{Reig} \& {Roche}}{{Reig} \&
  {Roche}}{1999}]{reig1999}
{Reig} P.,  {Roche} P.,  1999, \mn@doi [\mnras]
  {10.1046/j.1365-8711.1999.02463.x}, \href
  {http://adsabs.harvard.edu/abs/1999MNRAS.306...95R} {306, 95}

\bibitem[\protect\citeauthoryear{{Reig}, {Negueruela}, {Buckley}, {Coe},
  {Fabregat}  \& {Haigh}}{{Reig} et~al.}{2001a}]{reig2000}
{Reig} P.,  {Negueruela} I.,  {Buckley} D.~A.~H.,  {Coe} M.~J.,  {Fabregat} J.,
    {Haigh} N.~J.,  2001a, \mn@doi [\aap] {10.1051/0004-6361:20000238}, \href
  {http://adsabs.harvard.edu/abs/2001A%26A...367..266R} {367, 266}

\bibitem[\protect\citeauthoryear{{Reig}, {Negueruela}, {Buckley}, {Coe},
  {Fabregat}  \& {Haigh}}{{Reig} et~al.}{2001b}]{reig2001}
{Reig} P.,  {Negueruela} I.,  {Buckley} D.~A.~H.,  {Coe} M.~J.,  {Fabregat} J.,
    {Haigh} N.~J.,  2001b, \mn@doi [\aap] {10.1051/0004-6361:20000238}, \href
  {https://ui.adsabs.harvard.edu/\#abs/2001A&A...367..266R} {367, 266}

\bibitem[\protect\citeauthoryear{{Reig}, {Doroshenko}  \& {Zezas}}{{Reig}
  et~al.}{2014}]{Reig14}
{Reig} P.,  {Doroshenko} V.,   {Zezas} A.,  2014, \mn@doi [\mnras]
  {10.1093/mnras/stu1840}, \href
  {http://adsabs.harvard.edu/abs/2014MNRAS.445.1314R} {445, 1314}

\bibitem[\protect\citeauthoryear{{Rivinius}, {Carciofi}  \&
  {Martayan}}{{Rivinius} et~al.}{2013}]{rivinius2013}
{Rivinius} T.,  {Carciofi} A.~C.,   {Martayan} C.,  2013, \mn@doi [Astronomy
  and Astrophysics Review] {10.1007/s00159-013-0069-0}, \href
  {https://ui.adsabs.harvard.edu/\#abs/2013A&ARv..21...69R} {21, 69}

\bibitem[\protect\citeauthoryear{{Romanova}, {Ustyugova}, {Koldoba}  \&
  {Lovelace}}{{Romanova} et~al.}{2004}]{romanova2004}
{Romanova} M.~M.,  {Ustyugova} G.~V.,  {Koldoba} A.~V.,   {Lovelace} R.~V.~E.,
  2004, \mn@doi [\apj] {10.1086/426586}, \href
  {https://ui.adsabs.harvard.edu/#abs/2004ApJ...616L.151R} {616, L151}

\bibitem[\protect\citeauthoryear{{Rouco Escorial}, {van den Eijnden}  \&
  {Wijnands}}{{Rouco Escorial} et~al.}{2018}]{rouco2018}
{Rouco Escorial} A.,  {van den Eijnden} J.,   {Wijnands} R.,  2018, \mn@doi
  [\aap] {10.1051/0004-6361/201834572}, \href
  {https://ui.adsabs.harvard.edu/abs/2018A&A...620L..13R} {620, L13}

\bibitem[\protect\citeauthoryear{{Rutledge}, {Bildsten}, {Brown}, {Pavlov},
  {Zavlin}  \& {Ushomirsky}}{{Rutledge} et~al.}{2002}]{Rutledge02c}
{Rutledge} R.~E.,  {Bildsten} L.,  {Brown} E.~F.,  {Pavlov} G.~G.,  {Zavlin}
  V.~E.,   {Ushomirsky} G.,  2002, \mn@doi [\apj] {10.1086/342745}, \href
  {http://adsabs.harvard.edu/abs/2002ApJ...580..413R} {580, 413}

\bibitem[\protect\citeauthoryear{{Rutledge}, {Bildsten}, {Brown},
  {Chakrabarty}, {Pavlov}  \& {Zavlin}}{{Rutledge} et~al.}{2007}]{Rutledge07}
{Rutledge} R.~E.,  {Bildsten} L.,  {Brown} E.~F.,  {Chakrabarty} D.,  {Pavlov}
  G.~G.,   {Zavlin} V.~E.,  2007, \mn@doi [\apj] {10.1086/510183}, \href
  {http://adsabs.harvard.edu/abs/2007ApJ...658..514R} {658, 514}

\bibitem[\protect\citeauthoryear{{Shternin}, {Yakovlev}, {Haensel}  \&
  {Potekhin}}{{Shternin} et~al.}{2007}]{Shternin07}
{Shternin} P.~S.,  {Yakovlev} D.~G.,  {Haensel} P.,   {Potekhin} A.~Y.,  2007,
  \mn@doi [\mnras] {10.1111/j.1745-3933.2007.00386.x}, \href
  {http://adsabs.harvard.edu/abs/2007MNRAS.382L..43S} {382, L43}

\bibitem[\protect\citeauthoryear{{Shternin}, {Baldo}  \& {Haensel}}{{Shternin}
  et~al.}{2018}]{shternin2018}
{Shternin} P.~S.,  {Baldo} M.,   {Haensel} P.,  2018, \mn@doi [Physics Letters
  B] {10.1016/j.physletb.2018.09.035}, \href
  {https://ui.adsabs.harvard.edu/abs/2018PhLB..786...28S} {786, 28}

\bibitem[\protect\citeauthoryear{{Stella}, {White}  \& {Rosner}}{{Stella}
  et~al.}{1986}]{stella86}
{Stella} L.,  {White} N.~E.,   {Rosner} R.,  1986, \mn@doi [\apj]
  {10.1086/164538}, \href {http://adsabs.harvard.edu/abs/1986ApJ...308..669S}
  {308, 669}

\bibitem[\protect\citeauthoryear{{Tsygankov}, {Lutovinov}, {Doroshenko},
  {Mushtukov}, {Suleimanov}  \& {Poutanen}}{{Tsygankov}
  et~al.}{2016}]{tsygankov2016}
{Tsygankov} S.~S.,  {Lutovinov} A.~A.,  {Doroshenko} V.,  {Mushtukov} A.~A.,
  {Suleimanov} V.,   {Poutanen} J.,  2016, \mn@doi [\aap]
  {10.1051/0004-6361/201628236}, \href
  {http://adsabs.harvard.edu/abs/2016A%26A...593A..16T} {593, A16}

\bibitem[\protect\citeauthoryear{{Tsygankov}, {Wijnands}, {Lutovinov},
  {Degenaar}  \& {Poutanen}}{{Tsygankov} et~al.}{2017a}]{tsygankov2017a}
{Tsygankov} S.~S.,  {Wijnands} R.,  {Lutovinov} A.~A.,  {Degenaar} N.,
  {Poutanen} J.,  2017a, \mn@doi [\mnras] {10.1093/mnras/stx1255}, \href
  {http://adsabs.harvard.edu/abs/2017MNRAS.470..126T} {470, 126}

\bibitem[\protect\citeauthoryear{{Tsygankov}, {Mushtukov}, {Suleimanov},
  {Doroshenko}, {Abolmasov}, {Lutovinov}  \& {Poutanen}}{{Tsygankov}
  et~al.}{2017b}]{tsygankov2017b}
{Tsygankov} S.~S.,  {Mushtukov} A.~A.,  {Suleimanov} V.~F.,  {Doroshenko} V.,
  {Abolmasov} P.~K.,  {Lutovinov} A.~A.,   {Poutanen} J.,  2017b, \mn@doi
  [\aap] {10.1051/0004-6361/201630248}, \href
  {http://adsabs.harvard.edu/abs/2017A%26A...608A..17T} {608, A17}

\bibitem[\protect\citeauthoryear{{Wijnands}, {Degenaar}, {Armas Padilla},
  {Altamirano}, {Cavecchi}, {Linares}, {Bahramian}  \& {Heinke}}{{Wijnands}
  et~al.}{2015}]{wijnands2015}
{Wijnands} R.,  {Degenaar} N.,  {Armas Padilla} M.,  {Altamirano} D.,
  {Cavecchi} Y.,  {Linares} M.,  {Bahramian} A.,   {Heinke} C.~O.,  2015,
  \mn@doi [\mnras] {10.1093/mnras/stv1974}, \href
  {https://ui.adsabs.harvard.edu/\#abs/2015MNRAS.454.1371W} {454, 1371}

\bibitem[\protect\citeauthoryear{{Wijnands}, {Degenaar}  \& {Page}}{{Wijnands}
  et~al.}{2017}]{wijnands2017}
{Wijnands} R.,  {Degenaar} N.,   {Page} D.,  2017, \mn@doi [Journal of
  Astrophysics and Astronomy] {10.1007/s12036-017-9466-5}, \href
  {https://ui.adsabs.harvard.edu/\#abs/2017JApA...38...49W} {38, 49}

\bibitem[\protect\citeauthoryear{{Wilms}, {Allen}  \& {McCray}}{{Wilms}
  et~al.}{2000}]{wilms2000}
{Wilms} J.,  {Allen} A.,   {McCray} R.,  2000, \mn@doi [\apj] {10.1086/317016},
  \href {http://adsabs.harvard.edu/abs/2000ApJ...542..914W} {542, 914}

\makeatother
\end{thebibliography}








\bsp	
\label{lastpage}
\end{document}